\documentclass[a4paper,fleqn,usenatbib]{mnras}

\usepackage{newtxtext,newtxmath}
\usepackage[T1]{fontenc}
\usepackage{ae,aecompl}

\usepackage{graphicx}		  % Including figure files
\usepackage{amsmath}		% Advanced maths commands
\usepackage{amssymb}		% Extra maths symbols

\usepackage{siunitx}			          %SI unites for micro meter
\usepackage{cleveref}					 %use cref for sections
\crefformat{section}{\S#2#1#3}	        % make section § symbol
\usepackage[table]{xcolor}
\usepackage{import}						% import appendix in separate file
\usepackage{makecell}					% line-break in table column
\usepackage{enumerate}

\definecolor{lavendergray}{rgb}{0.77, 0.76, 0.82}
\definecolor{cadetgray}{rgb}{0.57, 0.64, 0.69}
\definecolor{gray}{rgb}{0.5, 0.5, 0.5}
\definecolor{debianred}{rgb}{0.84, 0.04, 0.33}
\definecolor{ballblue}{rgb}{0.13, 0.67, 0.8}
\definecolor{frenchblue}{rgb}{0.0, 0.45, 0.73}
\definecolor{indigo}{rgb}{0.0, 0.25, 0.42}

\title[Fundamental differences between red and blue quasars]{Fundamental differences in the radio properties of red and blue quasars: evolution strongly favoured over orientation}
\author[L. Klindt]{
	L. Klindt,$^{1}$\thanks{lizelke.klindt@durham.ac.uk}
	D. M. Alexander,$^{1}$
	D. J. Rosario,$^{1}$ 
	E. Lusso,$^{1}$
	\& S. Fotopoulou$^{1}$
	\\
	$^{1}$Centre for Extragalactic Astronomy, Department of Physics, Durham University, Durham DH1 3LE, UK\\
}

\date{Accepted XXX. Received YYY; in original form ZZZ}

\pubyear{2019}

\begin{document}
	\label{firstpage}
	\pagerange{\pageref{firstpage}--\pageref{lastpage}}
	\maketitle
	
	%%%%%%%%%%%%%%%%%%%%%%%%%%%%%%%%%%%%%%%%%%%%%%%%%%
	%%%%%%%%%%%%%%%%% ABSTRACT %%%%%%%%%%%%%%%%%%
	\begin{abstract}
		A minority of the optically selected quasar population are red at optical wavelengths due to the presence of dust along the line-of-sight. 
		A key focus of many red quasar studies is to understand their relationship with the overall quasar population: are they blue quasars observed at a (slight) inclination angle or do they represent a transitional phase in the evolution of quasars?
		Identifying fundamental differences between red and blue quasars is key to discriminate between these two paradigms. 
		To robustly explore this, we have uniformly selected quasars from the Sloan Digital Sky Survey with mid-infrared counterparts, carefully controlling for luminosity and redshift effects. We take a novel approach to distinguish between colour-selected quasars in the redshift range of $0.2 < z < 2.4$ by constructing redshift-sensitive $g^* - i^*$ colour cuts. 
		From cross-matching this sample to the Faint Images of the Radio Sky at Twenty-centimeters (FIRST) survey, we have found a factor $\approx$\,3 larger fraction of radio-detected red quasars with respect to that of blue quasars. Through a visual inspection of the FIRST images and an assessment of the 
		radio luminosities (rest-frame ${L_{\rm 1.4~GHz}}$ and ${L_{\rm 1.4~GHz}}/{L_{\rm 6\mu m}}$) we find that the radio-detection excess for red quasars is primarily due to compact and radio-faint systems (around the radio quiet--radio loud threshold). We show that our results rule out orientation as the origin for the differences between red and blue quasars and argue that they provide broad agreement with an evolutionary model.
	\end{abstract}
	%No significant differences are found between red and blue quasars within the classical extended radio-loud systems. 
	\begin{keywords}
		galaxies: active -- galaxies: evolution -- galaxies: jets -- quasars: general -- quasars: supermassive black holes -- infrared: galaxies -- radio-continuum: galaxies.
	\end{keywords}

	%%%%%%%%%%%%%%%%%%%%%%%%%%%%%%%%%%%%%%%%%%%%%%%%%%
	%%%%%%%%%%%%%%%%% BODY OF PAPER %%%%%%%%%%%%%%%%%%
	%==============================================================================================================
	%==============================================================================================================
	\section{ Introduction}
	\label{sec:introduction}
	%Add references: \\
	%Richards+2001 (radio QSOs are red);\\
	%Ivezic+2002 (radio quasars are red);\\
	%Glikman+2012; \\
	%Mehdipour+2019 (Relation between winds and jets in radio-loud AGN) \\
	Quasars, often referred to as Quasi-Stellar Objects (QSOs), represent the most luminous subset of the overall population of Active Galactic Nuclei (AGN). 
	Their prodigious output (bolometric luminosities up to 10$^{47}$\,erg\,s$^{-1}$) is indicative of rapid accretion, at or near the Eddington limit, onto a supermassive black hole (BH; 10$^9$\,M$_\odot$) that resides at the heart of the host galaxy.
	Quasars radiate across the electromagnetic spectrum, from radio to X-rays. 
	%%% read review by Padovani and Dave (2017). %%%
	%Thermal emission from the inner accretion disk is observed at soft X-ray energies, while at hard X-rays soft disc photons are upscattered to higher energies by the corona through inverse Compton scattering. 
	The primary emission comes from the accretion onto the BH (in the form of an accretion disc) which produces thermal emission at ultraviolet (UV)--optical wavelengths. Surrounding the accretion disc is a geometrically and optically thick structure of cold molecular gas and dust (commonly referred to as the dusty ``torus'') which obscures a direct view of the accretion disc from some viewing angles. The dust in the ``torus'' is heated by the accretion disc and emits thermally at mid-infrared wavelengths (MIR;  5\,--\,40\,$\mu$m), dropping off steeply to the far-infrared band \citep[FIR; 40~--~500\,$\mu$m; e.g.][]{elvis1994,richards2006,netzer2007,mullaney2011}.
	In the radio waveband ($>$1~cm) emission produced by relativistic jets emerging from the vicinity of the black hole (BH) is sometimes detected; overall $\approx$~5--10\% of optically selected quasars are powerful in the radio waveband \citep[e.g.][]{condon2013}.
	% see Condon+2013
	
	A standard spectroscopic signature of a quasar is the presence of broad emission lines superimposed onto a blue power-law continuum. 
	However, the discovery of a subset of quasars with redder continuum emission (i.e.,\ red optical/IR colours) has challenged the conventional view. 
	Despite many studies in the literature \citep[e.g.][]{rieke1982, webster1995, benn1998, kim1999, francis2000, richards2003, glikman2004, glikman2007, glikman2012, urrutia2009, banerji2012, stern2012, assef2013, ross2015, lamassa2016, tsai2017, kim2018}, the nature of red quasars remains uncertain. 
	The majority of studies ascribe the red colours as due to dust-reddening of the accretion-disc emission \citep[e.g.][]{webster1995, wilkes2002, glikman2004, glikman2007, rose2013, kim2018}. Nonetheless, depending on the selection and luminosity of the quasar, the red colours can also be due to (1) an excess of flux at longer wavelengths either from a red synchrotron component or the ``contamination'' of starlight from the host galaxy \citep[e.g.][]{serjeant1996, benn1998, francis2000, whiting2001}, or (2) an intrinsically red continuum due to differences in the accretion disc and/or Eddington ratio when compared to normal quasars \citep[e.g.][]{richards2003,young2008}.
	%On the other hand, several studies have reported that obscuring dust is not the primary reddening process in their red quasar populations and other mechanisms such as host galaxy and synchrotron emission can play a role \citep[e.g.,][]{francis2000, whiting2001}. 
	%\citet{richards2003} for example showed that for a minority of red quasars the reddening of the continuum is perchance due to intrinsic drivers such as accretion flow and black hole (BH) mass.
	
	Ever since the first focused studies of red quasars over 20 years ago there has been debate over the relationship between red and blue quasars. If the red-blue quasar dichotomy is equivalent to that between nearby obscured and unobscured AGN then the different optical colours will be largely due to the orientation of an anisotropic obscuring structure \citep[i.e.,\ the dusty ``torus''; e.g.,][]{antonucci1993,urry1995}. 
	In this scenario red quasars will be more obscured by dust since they are viewed at inclinations closer to the equatorial plane of the dusty ``torus'' than blue quasars (i.e.,\ a red quasar is simply an inclined blue quasar whereby the observed viewing angle intersects a larger column of dust). 
	
	Alternatively, a competing paradigm postulates that red and blue quasars are related within an evolutionary sequence that connects dust-obscured star formation (SF) with quasar activity through gas inflow and outflow/feedback \citep[e.g.,][]{sanders1988,hopkins2008,alexander2012}.
	With this model the rare red quasar population represents a brief transitional phase \citep[a few tens of Myr;][]{hopkins2008} between the dust-obscured SF and the blue quasar phase during which winds/jets drive away the obscuring dust, ultimately shutting down the SF and revealing an unobscured blue quasar.
	
	Distinguishing between these competing scenarios is a major focus of red quasar studies. On the basis of emission line kinematics and X-ray analysis, some studies favour the orientation scenario to explain the nature of red quasars \citep[e.g.,][]{wilkes2002, rose2013}. Conversely, through a slew of different selection approaches (e.g.,\ optical; near-infrared (NIR); radio) and multi-wavelength analyses, other studies have presented evidence that many red quasars show the properties expected for the brief evolutionary phase \citep[e.g.,][]{urrutia2008,banerji2012,banerji2017,glikman2012,glikman2015}: major-merger driven gas inflows, dust-obscured SF, and energetic outflows. Irregardless of the fundamental nature of red quasars, all of these studies agree that the origin of the red quasar colours is due to the obscuration of a blue quasar continuum by dust; however, the location of the obscuring dust (i.e.,\ in the nucleus versus the host galaxy) and its origin (i.e.,\ a consequence of the inclination of the dusty ``torus'' versus the aftermath of a galaxy major merger event) remains uncertain.
	
	%More conclusively is that the aforementioned studies agree that obscuring dust acts as the reddening mechanism, possibly located at %different distances from the nuclear region, for example at the outskirts of the torus or on larger scales across the host galaxy disc. 
	%However, their findings diverge from one another in terms of the dust obscuration emergence in red quasars; is the dust an aftermath of %major-mergers or do we merely view red quasars through a grazing angle? 
	
	A major challenge in comparing between different red quasar studies is the broad range of selection approaches adopted \citep[e.g.][]{webster1995, cutri2001, gregg2002, glikman2007, banerji2012, tsai2017}; e.g.,\ optical--NIR--MIR colour criteria; point-source morphologies; bright radio emission. Furthermore, most red quasar studies do not uniformly select a normal blue quasar sample to provide a reliable ``control" to demonstrate that any observed differences (e.g.,\ in redshift, luminosity, BH mass, Eddington ratio) are specific to the red quasar population rather than being a consequence of the different selection approaches. To robustly demonstrate that there are fundamental differences between red and blue quasars which cannot be attributed just to orientation, the quasar samples must be selected in a uniform manner, carefully controlling for any additional selection effects.
	
	In this work we have used the Sloan Digital Sky Survey \citep[SDSS;][]{york2000} to undertake a uniform selection of red and blue quasars to search for fundamental differences, and to hence distinguish between the evolutionary and orientation scenarios. 
	In \cref{sec:sample selection} we outline the multi-wavelength data we used to construct a quasar parent sample which is uniform in selection and unbiased in the radio waveband. We also define our red and blue quasar samples and constrain the amount of dust obscuration required to produce the optical colours of the red quasars. Based on our findings we construct luminosity-redshift matched subsamples which we use throughout the paper as a comparison to the full colour-selected quasar subsamples.
	In \cref{subsec:radio-detection fraction result} we explore the radio properties of red and blue quasars and find that red quasars show an enhanced radio-detection fraction in comparison to blue quasars across all redshifts. In \cref{subsec:radio morphologies} we investigate the radio morphologies to determine which morphological structures are associated with the surfeit of radio-detected red quasars, and in \cref{subsec:radio luminosities} we explore the relation between the radio-detection fraction and the radio luminosity (${L_{\rm 1.4~GHz}}$ and ${L_{\rm 1.4~GHz}}/{L_{\rm 6\mu m}}$) of red quasars. Overall we find a significant enhancement in the detection of compact and faint radio sources in the red quasar population, a result that becomes stronger towards radio-quiet quasars.
	These results strongly argue against a simple orientation model but are in broad agreement with an evolutionary model, which we discuss in \cref{sec:discussion}. 
	In this work we adopted a concordance flat $\Lambda$-cosmology with $H_0$ = 70\,km\,s$^{-1}$\,Mpc$^{-1}$ , $\Omega_{M}$ = 0.3, and $\Omega_\Lambda$ = 0.7.

	%==============================================================================================================
	%==============================================================================================================
	\section{Data sets and quasar sample definition}
	\label{sec:sample selection}
\begin{figure*}
		\centering
		\includegraphics[width=38pc]{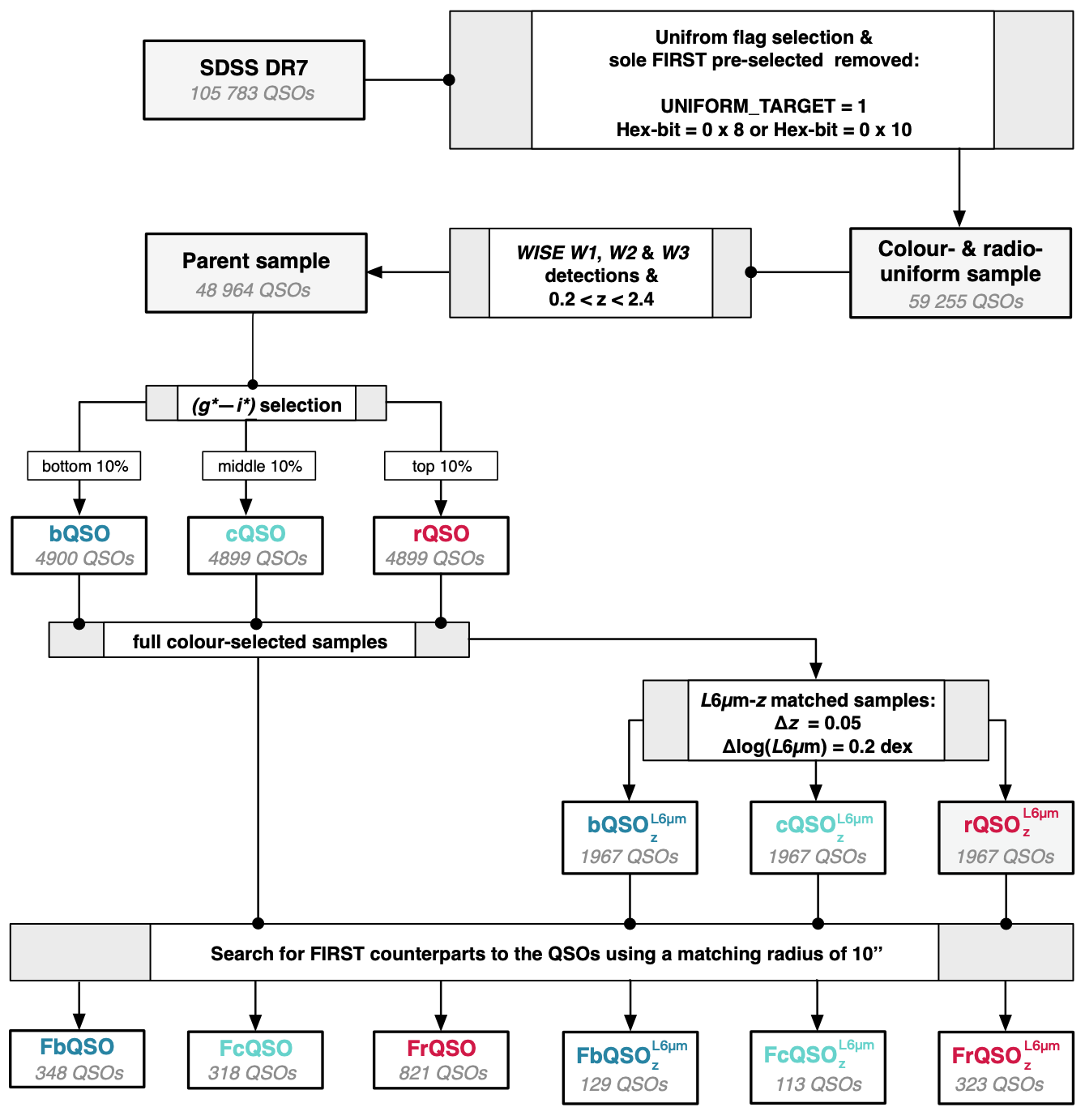}		
		\caption{A schematic diagram of our selection process. We started with quasars from \citet{schneider2010} and selected those with the uniform selection flag and excluded sources which were solely targeted for spectroscopic followup due to their FIRST pre-selection; i.e.,\ sources which were selected only due to having a FIRST counterpart and did not satisfy any of the other selection criteria outlined in \citet{richards2002}. To construct a parent sample we included sources at redshifts of $0.2 < z < 2.4$ with {\it WISE} detections in bands $W1$, $W2$ and $W3$ (SNR $> 2$), and with bolometric luminosity and BH mass measurements in \citet{shen2011}. The parent sample was then equally subdivided into the bottom, middle and top 10\% of the redshift dependent $g^* - i^*$ distributions representing our full colour-selected samples termed as bQSOs, cQSOs and rQSOs, respectively (see Figure~\ref{fig:gi_z_distribution}). Additionally, the bQSOs, cQSOs and rQSOs were matched in 6\,$\mu$m luminosity and redshift to construct $L_{\rm 6 \mu m}-z$ matched colour-selected subsamples. Following the colour selection and luminosity-redshift matching, we searched for FIRST radio counterparts within a 10$''$ search radius; these samples represent FIRST-detected quasars termed as FbQSO, FcQSOs and FrQSOs for the FIRST-detected bQSOs, cQSOs and rQSOs, respectively. The source statistics for the different subsamples split into redshift bins are tabulated in Table~\ref{tab:source_stats}.}
		\label{fig:flowchart}	
	\end{figure*}

	In this study we explore the physical properties of red quasars at $0.2 < z < 2.4$ to understand whether they are intrinsically different to the overall quasar population at the same epochs. 
	The multi-wavelength data and catalogues we used to select red and blue quasars are highlighted in \cref{subsec:data} and the details of our careful selection criteria are described in \cref{subsec:sample definition}; Figure~\ref{fig:flowchart} presents a flowchart that summarizes the different steps that are taken in defining our full colour-selected subsamples.
	In \cref{subsec:sample properties} we utilise the MIR emission as a robust tracer of the nuclear power of quasars and define a $L_{\rm 6 \mu m}-z$ matched sample to minimise luminosity and redshift effects on our results.
	%..............................................................................................................
	%..............................................................................................................
	
	%==============================================================================================================
	\subsection{Multi-wavelength data and catalogues}
	\label{subsec:data}
	
	Our quasar selection is based on the SDSS DR7 Quasar Catalogue \citep[][]{schneider2010}. We utilise this catalogue in combination with MIR data from the Wide-field Infrared Survey Explorer \citep[{\it WISE};][]{wright2010} and radio data from the Faint Images of the Radio Sky at Twenty-Centimeters \citep[FIRST;][]{becker1995,becker2012,helfand2015} and the NRAO VLA Sky Survey \citep[NVSS;][]{condon1998} to refine our selection approach and quasar analyses.
	
	%++++++++++++++++++++++++++++++++++++++++++++++++++++++++++++++++++++++++++++++++++++++++++++++++++++++++++++++
	\subsubsection{Optical data: the SDSS DR7 Quasar Catalogue} 
	\label{subsubsec:sdss data}
	The SDSS DR7 Quasar Catalogue \citep[][hereafter S10]{schneider2010} consists of 105,783 spectroscopically confirmed quasars with luminosities of $M_i < -22.0$ out to $z = 5.48$ which exhibit at least one emission line with a full width at half-maximum (FWHM) $> 1000$\,km\,s$^{-1}$ or exhibit broad absorption line features. We briefly describe the construction of the quasar catalogue below.
	
	The quasar selection algorithm described in \citet{richards2002} was used to select primary candidate quasars for spectroscopic followup \citep[see also][]{richards2001,stoughton2002,schneider2005,vandenberk2005}. This algorithm distinguishes between quasars and the much more numerous stars and galaxies by (1) using their nonstellar colours obtained from the $u^*$\,(3543\,\AA{}), $g^*$\,(4770\,\AA{}), $r^*$\,(6231\,\AA{}), $i^*$\,(7625\,\AA{}) and $z^*$\,(9134\,\AA{}) broadband photometry and (2) searching for counterparts in the FIRST survey to all unresolved objects brighter than $i^*_{\rm dered}$ = 19.1.\footnote{Hereafter all SDSS PSF magnitudes used in this paper refer to the Galactic-extinction corrected magnitudes provided in \citet{shen2011}.} The target colour selection is sensitive to red quasars \citep[e.g.,\ quasars with $E(B-V)$ = 0.1 have a high probability of being selected; see \S2.2 of][]{richards2003}, although the most reddened quasars will be missed.
	In addition, non-quasar selection algorithms were used to supplement the primary quasar selection, including (1) objects with \textit{ROSAT} All-sky Survey \citep[RASS,][]{voges1999,voges2000} counterparts, (2) objects targeted as members of certain stellar populations (e.g.,\ \textit{F} stars and Main-sequence turnoff stars) but whose spectra showed them to be quasars, and (3) serendipitous objects \citep[FIRST matches or objects with peculiar colours; see][for further details on the SDSS selection algorithms]{stoughton2002,anderson2003,vandenberk2005,richards2006_sdss,shen2007,schneider2010,shen2011}. 
	These candidates were only assigned fibres once the main samples of galaxies, Luminous Red Galaxies (LRGs) and quasars were tiled. Hence, these samples are incomplete and in combination with the inclusion of the ``Special Plates" in the Galactic cap \citep[e.g.,\ Stripe 82][]{stoughton2002,adelman-mccarthy2006}, were designed to explore the limits of the primary selection algorithm, to go deeper and to target objects with atypical colours.
	
	The selection of quasars via earlier versions of the algorithm or from the Special Plates introduced a non-uniformity in the selection of the S10 quasar sample and, therefore, approximately half of these objects are not suitable for statistical analyses \citep{shen2011, kratzer2015, park2015}.
	However, S10 identify targets that satisfy {\it a posteriori} the \citet{richards2002} selection algorithm, indicated with the uniform flag \citep[{\sc uniform\_target} = 1; see][]{shen2011,park2015,kratzer2015}, which provides a statistically reliable sample of 59,514 quasars up to $z = 5.48$. Of these, only 259 quasars were targeted uniquely by FIRST pre-selection.
	
	%++++++++++++++++++++++++++++++++++++++++++++++++++++++++++++++++++++++++++++++++++++++++++++++++++++++++++++++
	\subsubsection{Radio data: searching for FIRST counterparts} 
	\label{subsubsec:first data}
	The VLA Faint Images of the Radio Sky at Twenty-Centimeters \citep[FIRST;][]{becker1995,becker2012,helfand2015} is a 1.4~GHz radio survey that observed $\approx$\,10,000~deg$^2$ of the SDSS region at a spatial resolution of 5$''$. %with a surface density $\sim$\,90~deg$^{-2}$ and
	The 5\,$\sigma$ source detection threshold of 1\,mJy makes it beneficial to detect quasars down to low radio luminosities; i.e.,\ including even radio-quiet AGN.
	The FIRST survey comprises 946,432 radio sources of which 30\% have a spectroscopic SDSS counterpart \citep[e.g.,][]{ivezic2002}.
	
	When cross-matching between different surveys there is a trade-off between completeness and the number of false associations. 
	Since the majority of the SDSS quasars are likely to be unresolved in FIRST, a high completeness and low random association can be achieved even when adopting a small search radius.
	Based on the analysis of \citet{lu2007}, we adopted a 10$''$ cross-matching radius.
	They showed that the false association rate within 10$''$ of the QSO position is only 0.2\%, with just 2\% of radio quasars in SDSS having radio structures that extend beyond 10$''$ and have an undetected radio core in FIRST.
	Therefore, we consider all FIRST radio sources with centroidal positions within 10$''$ of the quasar to be directly associated, and summed their integrated flux. 
	In practice only 6\% of all our radio-detected quasars have multiple FIRST counterparts treated in this fashion (see \cref{subsec:radio morphologies}).
	To explore our incompleteness to large radio sources we used additional data from the NVSS \citep[][]{condon1998} which has a beam size of 45$''$, significantly larger than FIRST and our cross-match radius (\cref{subsec:radio morphologies}).

	We calculated the 1.4\,GHz luminosities using the methodology described in \citet{alexander2003}, assuming a uniform radio spectral index of $\alpha = 0.5$ to compute a K-correction \citep[which is the division between steep and flat radio spectrum quasars; e.g.,][]{wall1975,kimball2008}.\footnote{We define the spectral index $\alpha$ as $f_\nu \propto \nu^{-\alpha}$.} 
	In this work we used the FIRST integrated flux ($F_{\rm int}$) to compute the radio luminosities and the FIRST peak flux ($F_{\rm peak}$) to establish whether a source is radio faint (i.e.,\ $F_{\rm peak} < 3$\,mJy).
	
	%++++++++++++++++++++++++++++++++++++++++++++++++++++++++++++++++++++++++++++++++++++++++++++++++++++++++++++++
	\subsubsection{Infrared data: searching for {\it WISE} counterparts} 
	\label{subsubsec:WISE data}
	To explore the dust properties of red quasars in comparison to blue quasars, we searched for MIR counterparts from the Wide-field Infrared Survey Explorer \citep[{\it WISE};][]{wright2010} which mapped the entire sky in four bands: $W1$ ($\lambda$~=~3.4\,$\mu$m; $PSF$~=~6.1$''$), $W2$ ($\lambda$~=~4.6\,$\mu$m; $PSF$~=~6.4$''$), $W3$ ($\lambda$~=~12\,$\mu$m; $PSF$~=~6.5$''$) and $W4$ ($\lambda$~=~22\,$\mu$m; $PSF$~=~12.0$''$).
	Using the Query Engine from the Infrared Science Archive at NASA/IPAC we matched the uniformly selected S10 quasars to the All-Sky WISE Source Catalogue (ALLWISE) within a 2.7$''$ radius. 
	This ensured a 99.5\% certainty that the optical source is matched to the correct MIR counterpart \citep{lake2012}. 
	We found a WISE counterpart for 58,137 uniformly selected S10 quasars.
	%\footnote{\citet{shen2011} provide a WISE counterpart to 37,850 S10 quasars. However,their 6$''$ radius likely includes a number of spurious matches.}.
	
	Since the MIR emission is a robust tracer of the reprocessed accretion disc emission from dust that is free of the effects of extinction, we determined the intrinsic luminosities of the quasars on the basis of the MIR fluxes after ensuring that the MIR emission is not significantly contaminated by non-AGN processes (see \cref{subsubsec: WISE properties}). 
	We computed the rest-frame $6\mu$m luminosity (L$_{\rm 6\mu m}$) by log-linear interpolation or extrapolation of the fluxes in the $W2$ and $W3$ bands, assuming that they are equivalent to monochromatic fluxes at the effective wavelength of the filters.   
	
	%==============================================================================================================
	\subsection{Distinguishing between red and blue quasars: full colour-selected samples} 
	\label{subsec:sample definition}
	A variety of different approaches have been adopted in the literature to select red quasars which makes it difficult to draw conclusions about their properties with respect to blue quasars. 
	Our main aim is to construct a carefully controlled and redshift-sensitive experiment in which both red and blue quasars are drawn from the same parent sample to allow for a systematic exploration of their multi-wavelength properties. 
	In \cref{subsubsec:parent sample} we construct a parent sample using the aforementioned catalogues/surveys and in \cref{subsubsec:colour selected samples} we define our red and blue quasar samples, referred to here as our full colour-selected samples.  
	
	%++++++++++++++++++++++++++++++++++++++++++++++++++++++++++++++++++++++++++++++++++++++++++++++++++++++++++++++
	\subsubsection{Defining the quasar parent sample}
	\label{subsubsec:parent sample}
	As mentioned in \cref{subsubsec:sdss data}, the SDSS quasar selection approach is complex which introduces some non-uniformity in the overall quasar sample. 
	We therefore sought a selection approach which is uniform and minimises radio biases introduced by the FIRST pre-selection of SDSS quasars.
	
	%\footnote{Approximately 5\% of the full S10 quasar catalogue were %FIRST pre-selected via matching unresolved optical sources to the %FIRST survey.}.      
	
	Figure~\ref{fig:flowchart} shows a flowchart that summarizes the different steps we took in defining our parent sample to then select red and blue quasars (\cref{subsubsec:colour selected samples}); the source statistics in each sample are reported in Table~\ref{tab:source_stats}.
	To construct a uniform parent sample we, firstly, selected S10 quasars with the uniform flag and then, using the hexadecimal bit value of the `BEST' flag \citep[see][]{stoughton2002,schneider2010}, we identified and removed sources pre-selected solely on the basis of their FIRST detection ({\sc bestflag} = 0\,$\times$\,8 for sources at high Galactic latitude and {\sc bestflag} = 0\,$\times$\,10 for sources at low Galactic latitude).
	%removed sole FIRST pre-selected sources to construct a colour- and radio-uniform sample, comprising of 59,255 quasars.  
	
	In addition to uniform selection, our sample is restricted to the redshift range $0.2 < z < 2.4$ and required a SNR $> 2$ in the $WISE$ $W1$, $W2$, and $W3$ bands to allow the estimation of $L_{\rm 6\mu m}$. We excluded 70 sources of the uniform sample which lacked bolometric luminosity measurements. 
	%Following the uniform selection we searched for a {\it WISE} counterpart to the optically selected quasars and restricted our sample to a redshift range of $0.2 < z < 2.4$ (see \cref{subsubsec:colour selected samples}). 
	%We further limited our sample by only including sources with a signal-to-noise ratio (SNR) $> 2$ in the $W1$, $W2$ and $W3$ bands, and excluded sources with missing data; i.e.,\ 70 quasars of the uniform sample which lacked bolometric luminosity measurements. 
	This yielded a final parent sample of 48,964 quasars that are uniform in selection and unbiased in the radio waveband within our selected redshift range.
	%--------------------------------------------------------------------------------------------------
	\bgroup
	\def\arraystretch{1.3}
	\begin{table}
		\begin{center}
			\small
			\caption{\label{tab:source_stats} Source statistics for the quasar samples in four different redshift bins. We present the number of sources for the uniformly selected parent sample, the full colour-selected samples (bQSOs, cQSOs and rQSOs), the $L_{\rm 6\mu m}-z$ matched colour samples (bQSO$^{L_{\rm 6\mu m}}_{z}$, cQSO$^{L_{\rm 6\mu m}}_{z}$ and rQSO$^{L_{\rm 6\mu m}}_{z}$) and their respective FIRST-detected subsamples. }
			\begin{tabular}[c]{lllllllccccccc}
				\hline
				\hline
				Sample & N$_{z1-z4}$ & N$_{z1}$ & N$_{z2}$   & N$_{z3}$   & N$_{z4}$  \\
				(1) & (2) & (3) & (4) & (5) & (6)  \\
				\hline
				\hline
				Parent & 48,964 & 5,402 & 6,021 & 18,286 & 19,255 \\
				\\
				\hline
				\multicolumn{6}{c}{Colour-selected QSO samples}\\
				\hline
				bQSOs & 4,900 & 535 & 613 & 1,822 & 1,923  \\
				cQSOs & 4,899 & 543 & 597 & 1,826 & 1,929\\
				rQSOs & 4,899 & 545 & 590 & 1,833 & 1,930\\
				\\
				FbQSOs & 348 & 62 & 71 & 110 & 105 \\
				FcQSOs & 318 & 52 & 48 & 121 & 97\\
				FrQSOs & 821 & 99 & 127 & 298 & 297\\
				\\
				\hline
				\multicolumn{6}{c}{Matched $L_{\rm 6\mu m}$--$z$ QSO samples} \\
				\hline
				bQSO$^{L_{\rm 6\mu m}}_{z}$ & 1,967 & 159 & 256 & 781 & 771 \\
				cQSO$^{L_{\rm 6\mu m}}_{z}$& 1,967 & 161 & 252 & 780 & 773\\
				rQSO$^{L_{\rm 6\mu m}}_{z}$ & 1,967 & 161 & 252 & 781 & 772\\
				\\
				FbQSO$^{L_{\rm 6\mu m}}_{z}$& 129 & 12 & 32 & 44 & 41 \\
				FcQSO$^{L_{\rm 6\mu m}}_{z}$ & 113 & 13 & 20 & 43 & 37\\
				FrQSO$^{L_{\rm 6\mu m}}_{z}$ & 323 & 39 & 51 & 121 & 112\\
				\\
				\hline
				\hline
			\end{tabular}
		\end{center}
		{\bf Notes.} (1): Target samples used in this study: Parent QSOs represents the uniformly selected S10 quasars with $W1-W3$ detections, b-, c-, rQSOs are the $g^* - i^*$ colour-selected samples and F-b-, c-, rQSOs are the radio-bright quasars matched to FIRST (see Figure~\ref{fig:gi_z_distribution}). (2)--(6): Source statistics for each sample within the respective redshift bins; i.e.,\ across the full $z$ range in our study, $0.2 < z_1< 0.5$, $0.5 < z_2 < 0.8$, $0.8 < z_3 < 1.5$ and $1.5 < z_4 < 2.4$.
	\end{table}
	\egroup
	%--------------------------------------------------------------------------------------------------

	%++++++++++++++++++++++++++++++++++++++++++++++++++++++++++++++++++++++++++++++++++++++++++++++++++++++++++++++
	\subsubsection{Defining the full colour-selected samples}
	\label{subsubsec:colour selected samples}
	To distinguish between red and blue quasars we produced $g^* - i^*$ colour distributions of our uniformly selected parent sample (\cref{subsubsec:WISE data}) as a function of redshift. 
	We define red quasars (rQSOs) and blue quasars (bQSOs) by selecting the respective reddest 10\% and the bluest 10\% of the colour distribution.
	Additionally we selected the 10\% of quasars around the median of the $g^* - i^*$ distribution to construct a control sample (cQSOs). 
	Hereafter we refer to all non-red quasars (including the bQSO and cQSO subsamples) as blue quasars.
	
\begin{figure*}
		\centering
		\includegraphics[width=40pc]{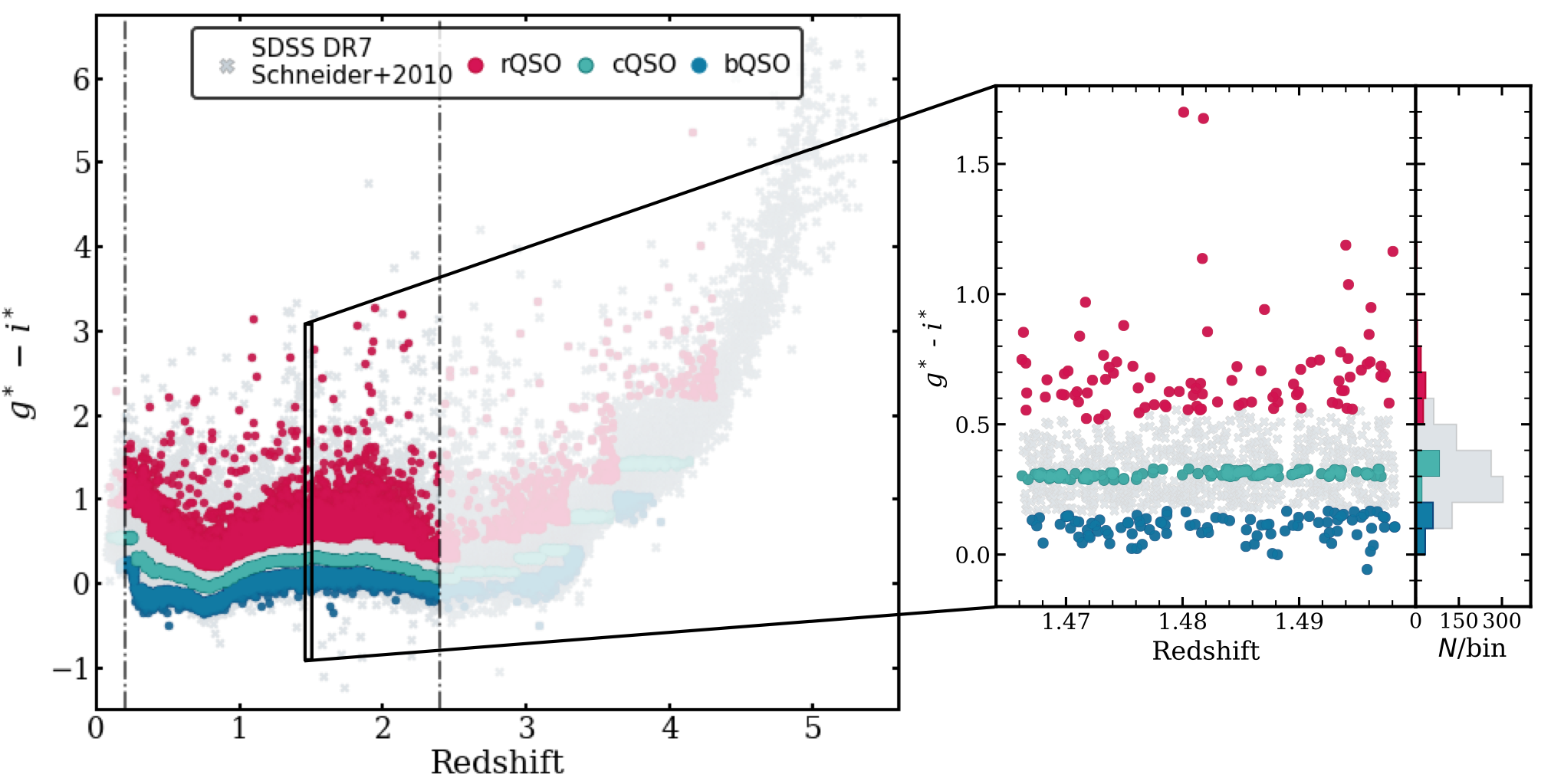}		
		\caption{The $g^* - i^*$ colours versus redshift for the bQSOs (blue circles), cQSOs (green circles) and rQSOs (red circles) explored in this study. Our selected quasars are superimposed on the distribution of S10 \citep[grey;][]{schneider2010}. Also indicated is the redshift range of our study ($z = 0.2-2.4$; dash-dot black line). {\it Zoom-in panel}: An illustration of the selection of the \mbox{bQSOs} (bottom 10\%), cQSOs (middle 10\%) and rQSOs (top 10\%) using the $g^* - i^*$ colours and redshift for an example bin of 1000 sources. The $g^* - i^*$ distribution is shown on the right. }
		\label{fig:gi_z_distribution}	
	\end{figure*}

	In Figure~\ref{fig:gi_z_distribution} we illustrate the selection of the three colour-selected quasar samples.
	In order to construct a rQSO selection which is sensitive to the redshift evolution of quasar SEDs we sorted the quasars by redshift and constructed the $g^* - i^*$ distributions in contiguous redshift bins consisting of 1000 sources as shown in the zoom-in panel in Figure~\ref{fig:gi_z_distribution}. 
	We restricted our analysis to $z \leq 2.4$ due to the low optical completeness of quasars in S10 at $2.5 < z < 3.0$. This is a consequence of the crossing between the stellar and quasar loci in the SDSS multicolour space \citep[see e.g.,][]{richards2002}. 
	%resulting in the followup of candidates which are identified based on other properties than their optical colours (i.e.,\ radio or X-ray detections).
	%..............................................................................................................
	%..............................................................................................................
	%%% see also Richards+03 where he explains why to use g-i.
	Even though the $u^* - z^*$ colour provides the broadest wavelength baseline for colour separation, the SDSS photometry in both of these bands is shallow and is more affected by atmospheric attenuation \mbox{\citep[e.g.,][]{ivezic2002}.}  
	Furthermore, the $u^*$-band is heavily affected by the Lyman break at $z \geq 1.9$ which motivates our use of the $g^*$-band (only influenced at higher redshifts, i.e.,\ $z \geq 2.5$).
	
\begin{figure}
		\centering
		\includegraphics[width=20pc]{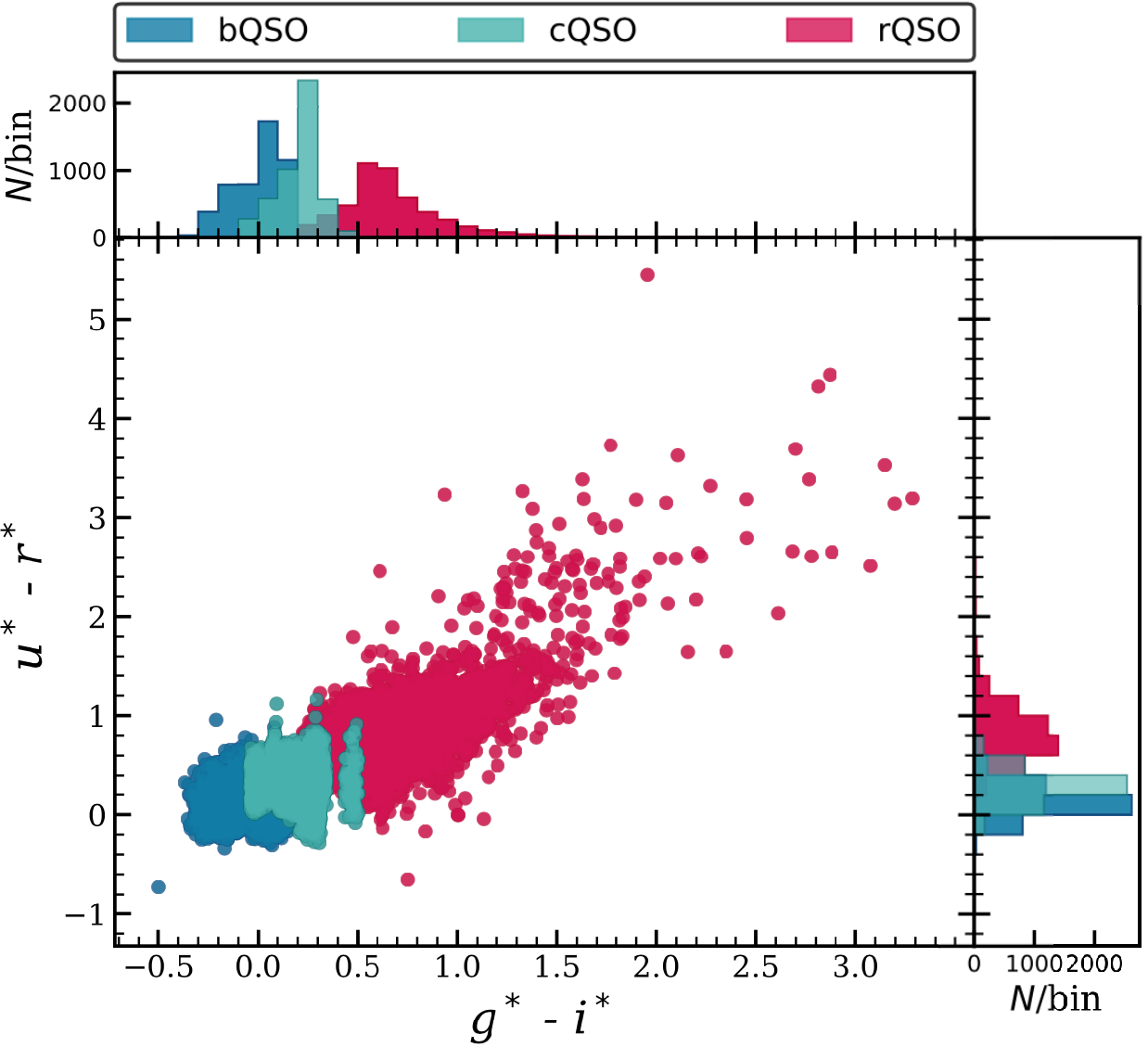}		
		\caption{A colour-colour diagram of the $u^* - r^*$ vs. $g^* - i^*$ colours for the full colour-selected samples; the bQSOs, cQSOs and rQSOs are indicated using the colour scheme in Figure~\ref{fig:gi_z_distribution} for redshifts in the range $0.2 < z < 2.4$. It is evident that the colours of red quasars are more broadly distributed in the colour-colour parameter space than more blue quasars, even in bands unrelated to our red quasar selection approach. Furthermore, our $g^* - i^*$ colour selection appears to be broadly consistent with other optical colours in selecting quasars, e.g.,\ $u^* - r^*$. A minority of the cQSOs have $g^* - i^* > 0.5$ which is not a failure of our approach but a normal consequence of our redshift-sensitive selection technique; see also Figure~\ref{fig:gi_z_distribution} where this is evident for quasars at $z < 0.3$.}
		\label{fig:sdss colour-colour}	
	\end{figure}

	Therefore, the $g^* - i^*$ colour was used to select red and blue quasars with the broadest possible wavelength range while optimising photometric depth \citep[see e.g.,][]{richards2003}.
	To verify that our colour selection does indeed reliably identify quasars with red and blue colours, and is not strongly influenced by broad emission line contamination in the $g^*$ or $i^*$ bands, we compared the $g^* - i^*$ to $u^* - r^*$ colours of the bQSOs, cQSOs and rQSOs as shown in Figure~\ref{fig:sdss colour-colour}. 
	We note that the rQSOs selected using $g^* - i^*$ are also red in $u^* - r^*$ when comparing to the bQSOs and cQSOs. 
	Approximately 2\% of the cQSOs have $g^* - i^* > 0.5$ (all of these sources are at $z < 0.3$; see Figure~\ref{fig:gi_z_distribution}).
	However, this is not a failure of our quasar selection approach and is a natural consequence of our redshift-sensitive selection technique. Overall, within the full colour-selected sample there are $\approx$~4900 quasars in each of the bQSO, cQSO, and rQSO sub samples, of which 348, 318, and 821 are radio detected, respectively.
	%..............................................................................................................
	%..............................................................................................................

	%++++++++++++++++++++++++++++++++++++++++++++++++++++++++++++++++++++++++++++++++++++++++++++++++++++++++++++++
	\subsubsection{Dust extinction in red quasars}
	\label{subsubsec: dust extinction}
	% see Richards+03  --> red and reddened quasars
	% also see Young+2008
	%Having explored the effects of the aforementioned mechanisms we now study the effects of dust based on previous studies.......
	%-- delta(g-i) 
	%-- dust is not the necessary ingredient for all of them
	%-- we do not have any close handle on the scale of the dust ---- transition into next section where dust can also be on large scales
	The most common explanation for the optical colours of red quasars is dust extinction, whether due to the inclination of a dusty torus resulting from an increase in dust along the line-of-sight or an obscuring dust envelope in which the young red quasar is embedded. We note that, due to the optical selection criteria, the SDSS will miss the most reddened quasars both because of their colours (red quasars overlay the stellar locus in most SDSS colour-colour diagrams) and because of the optical survey flux limit \citep[e.g.,][use NIR data and select SDSS optical drop outs to define their red quasar samples]{glikman2004,banerji2012}. 
	
\begin{figure}
		\centering
		\includegraphics[width=20pc]{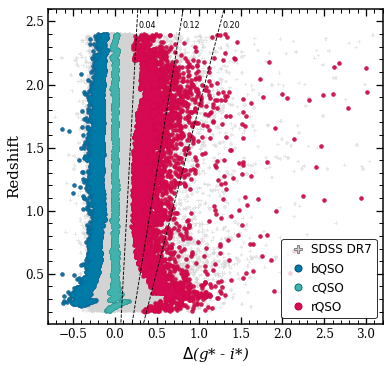}		
		\caption{Redshift vs. $\Delta (g^* - i^*)$ for the bQSOs, cQSOs and rQSOs using the colour scheme in Figure~\ref{fig:gi_z_distribution}. We illustrate the relative colours for our parent sample as well (faded grey points). The dashed lines are the dust-reddening for $E(B-V)$ = 0.04, 0.12 and 0.2 \citep[]{richards2003}, which correspond to $A_{\rm V}$ = 0.1, 0.3, and 0.5~mags for a typical quasar SED. The range in the relative colours of the rQSOs suggests that the majority require $A_{\rm V} \sim 0.1-0.5$\,mag to produce the observed optical colours.}
		\label{fig:delta_gi_fig}	
	\end{figure}

	In Figure~\ref{fig:delta_gi_fig} we visually examine the amount of dust reddening implied by our selection technique by plotting the $\Delta (g^* - i^*)$ colour as a function of redshift for the bQSOs, cQSOs and rQSOs. 
	The dashed lines denote the effect of SMC-type extinction \citep{prevot1984} as a function of redshift with $E(B-V)$ = 0.04, 0.12 and 0.2 \citep[e.g.,][]{richards2003} on the emission of a typical quasar. 
	Our selection of red quasars is broadly consistent with $E(B-V) > 0.04$ at $z > 0.8$ for a blue quasar SED.
	This corresponds to an equivalent hydrogen column density of $N_{\rm H} > $ 2.8 $\times$ $10^{20}$\,cm$^{-2}$ assuming an SMC-like dust-to-gas ($E(B-V)/N_{\rm H}$) ratio, comparable to that found towards the HII-regions in normal galaxies \citep{buchner2017}; we note that \citet{lamassa2016} and \citet{glikman2017} find that radio-selected red quasars have dust-to-gas ratios up-to an order of magnitude below the Galactic value, which would lead to higher hydrogen column densities by up-to an order of magnitude.
	The range in $\Delta (g^* - i^*)$ colours suggest that the majority of our rQSOs require $A_{\rm V} \sim 0.1-0.5$\,mag to produce the observed optical colours. By comparison,
	NIR based selection techniques are sensitive to selecting red quasars with a dust extinction of up to $A_{\rm V} \sim 1-6$\,mag \citep[e.g.][]{glikman2004,banerji2012}, hence the rQSOs in our study represent less extreme dust-reddened red quasars. 
	
	%Our rQSOs require $A_{\rm V} \sim 0.1-0.5$\,mag to produce the observed colours and, therefore, are not heavily obscured AGN, consistent with the presence of broad emission lines. 
	%--------------------------------------------------------------------------------------------------
	%--------------------------------------------------------------------------------------------------

	%==============================================================================================================
	\subsection{Defining the AGN power at rest-frame 6\,$\mu$m: luminosity-redshift matched samples} 
	\label{subsec:sample properties}
	To reliably infer any fundamental differences in the properties of the rQSOs from blue quasars (i.e.,\ the bQSOs and cQSOs) we must ensure that our results are not driven by differences in the bolometric luminosities of the quasars.
	\citet{shen2011} provide bolometric luminosities ($L_{\rm bol}$) for the S10 quasar sample, however, they are inferred from rest-frame UV-optical continuum measurements and have not been corrected for dust extinction. 
	Consequently, the $L_{\rm bol}$ values are likely to be significantly underestimated in the rQSOs, which will also lead to unreliable BH masses and Eddington ratios. We therefore calculate the bolometric luminosities for our quasar subsamples using the MIR {\it WISE} data, as described below, and utilise these to create luminosity-redshift matched samples.
	
	%++++++++++++++++++++++++++++++++++++++++++++++++++++++++++++++++++++++++++++++++++++++++++++++++++++++++++++++
	\subsubsection{MIR measurements: a more robust approach}
	\label{subsubsec: WISE properties}
\begin{figure*}
		\centering
		\includegraphics[width=35pc]{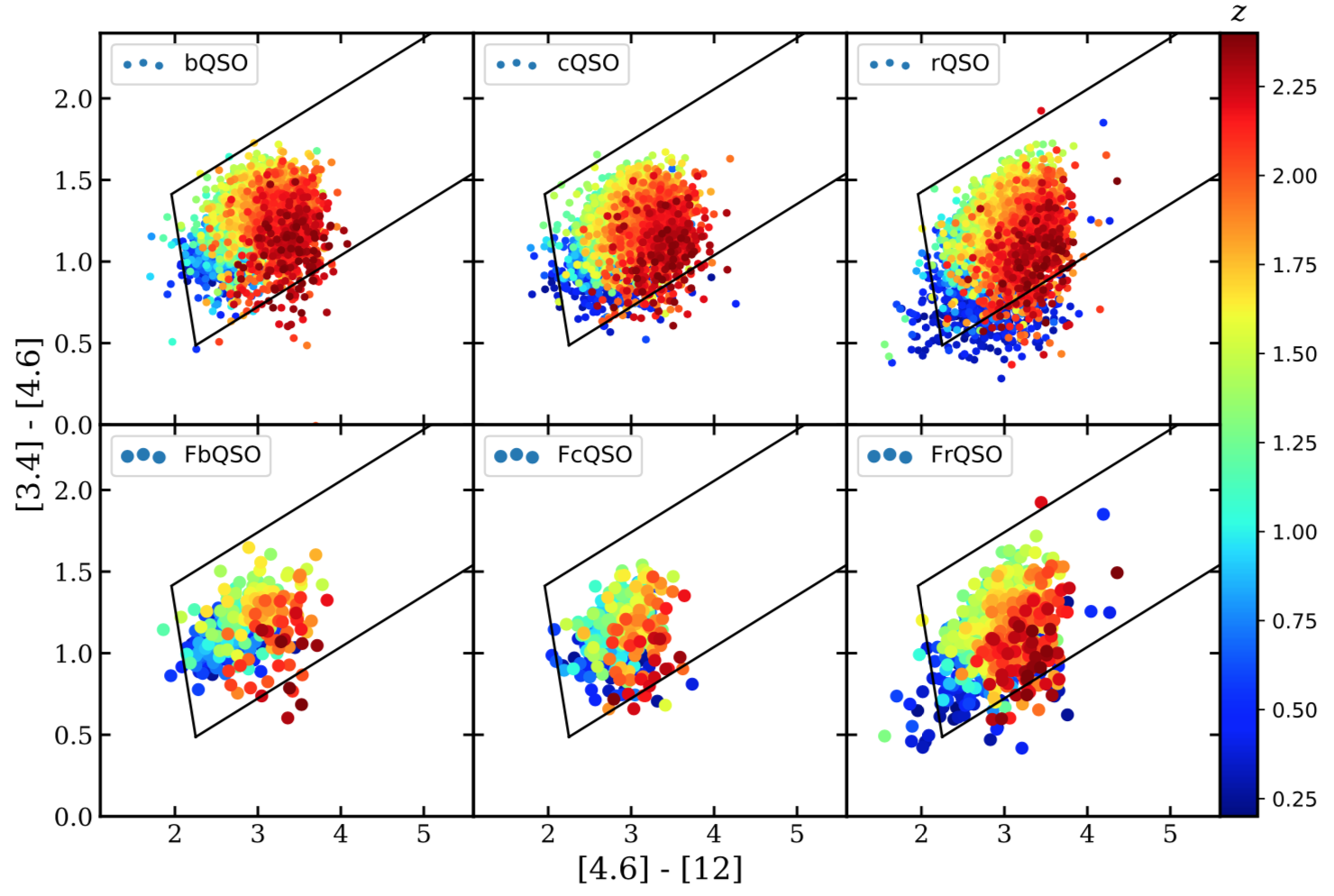}		
		\caption{A {\it WISE} colour-colour diagram for our optically selected quasars colour coded by redshift. The \citet{mateos2012} AGN wedge is indicated with black solid lines. The bQSOs, cQSOs and rQSOs are shown in the top panels from left to right, and the radio-detected FbQSOs, FcQSOs and FrQSOs are presented in the bottom panels from left to right, respectively. The majority of the quasars lie in the AGN wedge; see Table~\ref{tab:outside_wedge}.}
		\label{fig:WISE_wedge}	
	\end{figure*}

	%To discern whether the MIR emission is dominated by the AGN we make use of the WISE three-band AGN wedge introduced by \citet{mateos2012}, which defines a reliable MIR colour selection of luminous AGN in the WISE colour-colour diagram (called the "AGN wedge") and identifies AGN with red MIR power-law SEDs with a spectral index $\alpha \leq -0.3$.
	In order to accurately quantify the MIR emission from the quasar, it is imperative to verify that it is not contaminated by the host galaxy. Since the dusty AGN torus radiates predominantly at MIR wavelengths while star formation from the host galaxy peaks at FIR wavelengths, we would expect the MIR emission from our quasars to be dominated by the AGN. To verify this, we make use of the diagnostic $WISE$ three-band colour-colour diagram on which \citet{mateos2012} define a region (called the ``AGN wedge"; Figure~\ref{fig:WISE_wedge}) which identifies AGN with red MIR power-law SEDs with a spectral index $\alpha \leq -0.3$. 
	Host star-formation with a MIR luminosity $>10$\% of that of the AGN would systematically move QSOs out of the lower right of the wedge; however, this requires high star formation rate (SFR) levels (e.g.,\ $> 100$ M$_{\odot}$/yr at $z > 1$).

	In Figure~\ref{fig:WISE_wedge} we plot the WISE colours ($W1-W2$ vs $W2-W3$) of our colour-selected quasars and indicate the AGN wedge to show where AGN are expected to lie. 
	The bulk ($\sim$95-99\%) of the quasars lie within the wedge; however, a small fraction lie outside as reported in Table~\ref{tab:outside_wedge}.
	Most of the bQSO and cQSO outliers are higher redshift sources ($z~>~1.5$), while the majority of the rQSO outliers tend to be at lower redshifts ($z~<~1.5$); the same result is found for the FIRST-detected samples. 
	The colours of the sources outside of the wedge at low redshifts can potentially be attributed to dominant host galaxy emission in low-luminosity quasars. 
	However, since only a small fraction of the quasars lie outside of the AGN wedge we can confidently assert that $>90$\% of the MIR light in most of our quasars come from an AGN, and for these, we can reliably estimate their MIR AGN luminosities ($L_{\rm 6\mu m}$; see \cref{subsubsec:WISE data} for the details of the luminosity calculation).
	%--------------------------------------------------------------------------------------------------
	%--------------------------------------------------------------------------------------------------
	%-------------------------------------------------------------------------------------------------- 
	\begin{table}
		\begin{center}
			\small
			\caption{\label{tab:outside_wedge} The percentage of bQSOs, cQSOs and rQSOs which lie outside the \citet{mateos2012} AGN wedge across $0.2 < z < 2.4$. Tabulated are both the colour-selected SDSS samples, as well as the FIRST-detected QSOs.  }
			\begin{tabular}[]{ccccc}
				\hline
				\hline
				Sample & Nu. outside & \multicolumn{3}{c}{Percentage outside} \\
				& & $0.2 < z < 2.4$ &$z < 1.5 $ & $z > 1.5 $ \\
				\hline
				\hline
				\multicolumn{5}{c}{Colour-selected quasars} \\
				\hline
				bQSO & 56 & 1\% & 0.3\% & 0.8\% \\
				cQSO & 82 & 2\% & 0.8\% & 0.9\% \\
				rQSO & 305 & 6\% & 4\% & 2\% \\
				\hline
				\multicolumn{5}{c}{FIRST-detected quasars} \\
				\hline
				FbQSO &  8 & 2\% & 0.9\% & 1\% \\
				FcQSO &  9 & 3\% & 1\% & 2\% \\
				FrQSO &  43 & 5\% & 3\% & 2\% \\ 
				\hline
				\hline
			\end{tabular}
		\end{center}
	\end{table}
	%--------------------------------------------------------------------------------------------------
	
	%++++++++++++++++++++++++++++++++++++++++++++++++++++++++++++++++++++++++++++++++++++++++++++++++++++++++++++++
	%\subsubsection{The nuclear power in red quasars compared to blue quasars}
	%\label{subsubsec: L6um}
	
\begin{figure*}
		\centering
		\includegraphics[width=35pc]{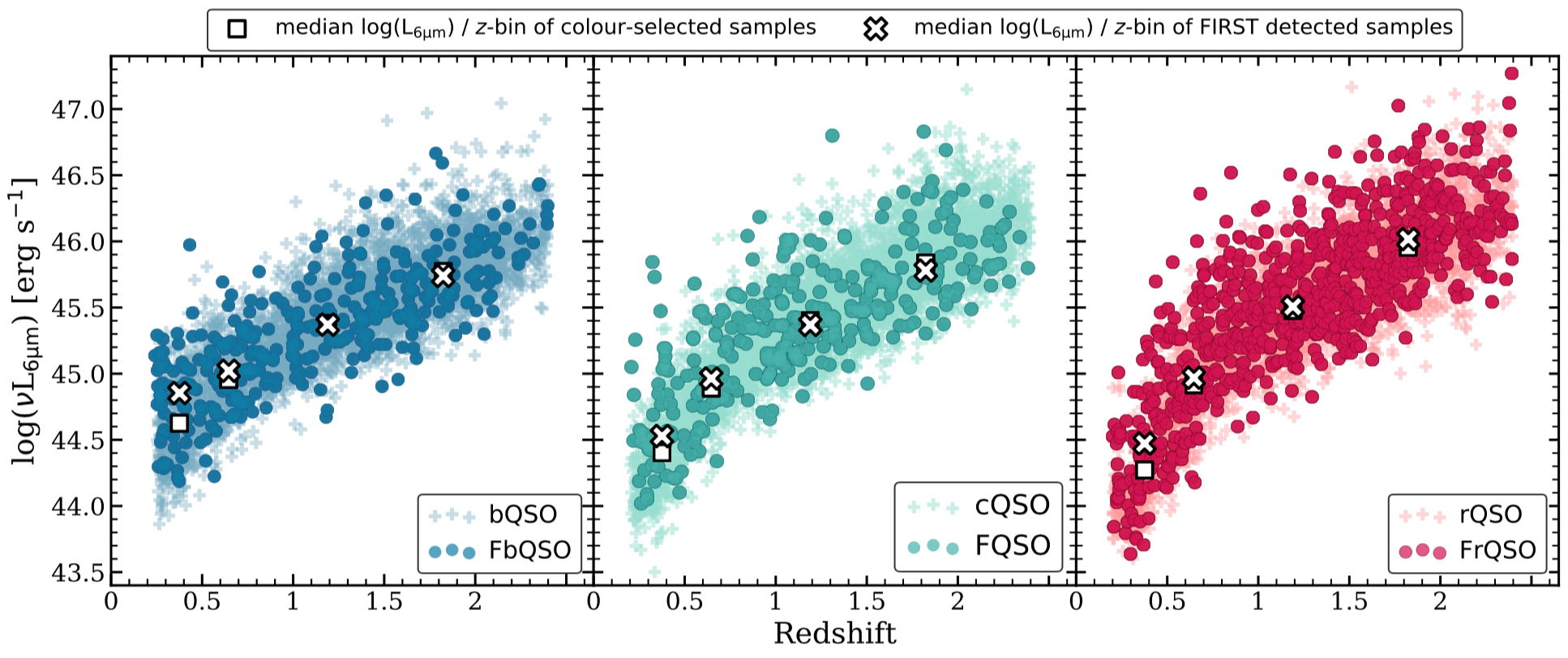}		
		\caption{Rest-frame 6\,$\mu m$ luminosity versus redshift for the bQSO, cQSO and rQSO quasars with the radio-detected FbQSOs, FcQSOs and FrQSOs overlaid. The median values for full colour-selected samples (X's) and FIRST-detected samples (squares) in each redshift bin are also plotted. }
		\label{fig:logL6um vs z}	
	\end{figure*}

	%The rest-frame 6\,$\mu$m luminosity (L$_{\rm 6 \mu m}$) is calculated by log-linear interpolation or extrapolation of the fluxes in the $W2$ and $W3$ bands, assuming that they are equivalent to monochromatic fluxes at the effective wavelength of the filters.
	In Figure~\ref{fig:logL6um vs z} we plot the $L_{\rm 6\,\mu m}$ as a function of redshift. 
	%(see also Figure~\ref{fig:l6um hist} for the distributions in the different $z$-bins). 
	The median $\log(L_{6 \mu m})$ of the colour-selected (square) and FIRST-detected (cross) quasars are also plotted in the respective $z$-bins. 
	%..............................................................................................................
	%..............................................................................................................
	We find lower median $\log(L_{6 \mu m})$ values for the rQSOs than the bQSOs at low redshifts ($z < 0.8$). This is also mirrored in the higher fraction of rQSO outliers in the AGN wedge at these redshifts and may be the consequence of host-galaxy dilution affecting the colours of the lower luminosity quasars at lower redshifts. 
	However, at higher redshifts ($0.8 < z < 2.4$) the opposite trend is seen where the rQSOs peak at higher $6\,\mu$m luminosities in comparison to the bQSOs and cQSOs (both in terms of higher median luminosities and larger maximum values). 
	These effects could be a consequence of the optical selection of SDSS quasars which would mean that rQSOs that are obscured in the optical/UV by dust will only satisfy the SDSS quasar selection if they are intrinsically more luminous.
	
\begin{figure*}
		\centering
		\begin{minipage}[c]{\textwidth}
			\centering
			\includegraphics[width=40pc]{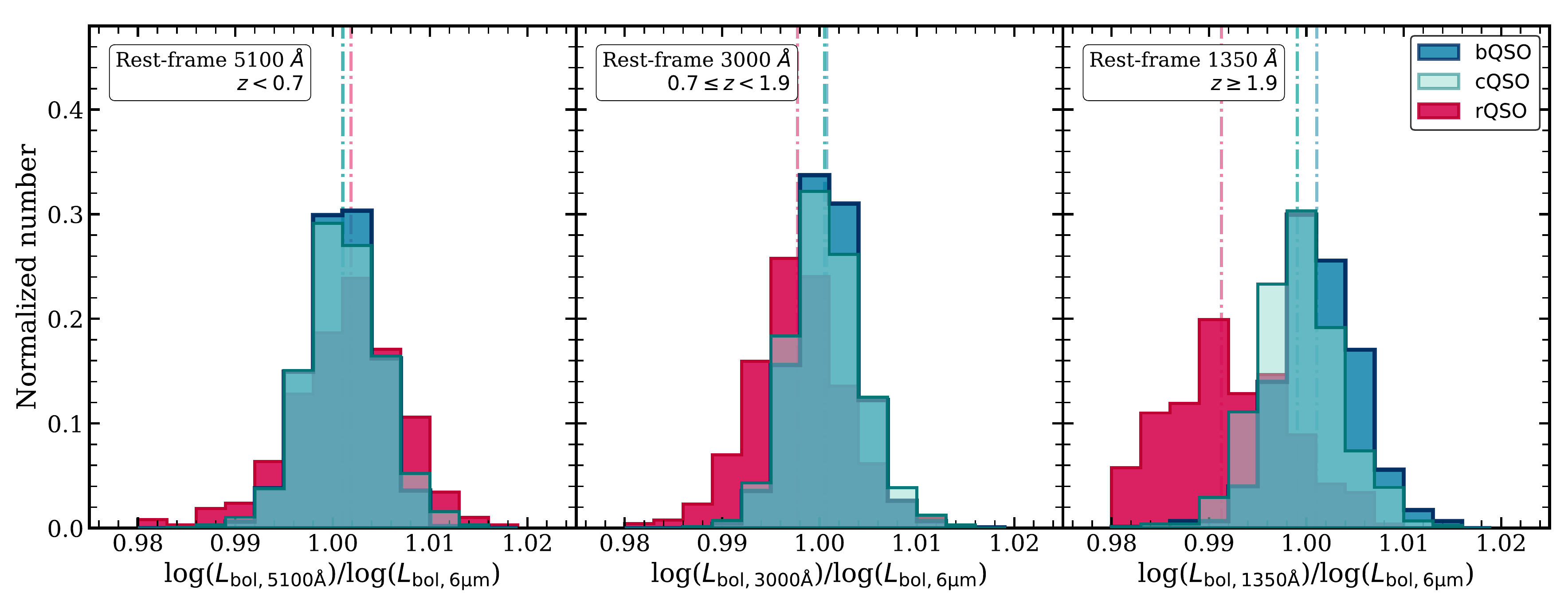}		
			\caption{Luminosity ratio distributions using the continuum luminosities at rest-frame 5100\,\AA{} (left, BC = 9.26), 3000\,\AA{} (middle, BC = 5.15) and 1350\,\AA{} (right, BC = 3.81) adopted from \citet{shen2011} and the inferred bolometric luminosity from the rest-frame $6\,\mu$m luminosity with a bolometric correction of BC $= 8$. The vertical dash-dot lines illustrate the median luminosity ratio for the respective radio-detected quasar subsamples.}
			\label{fig:L6mu vs Lbol fraction}
		\end{minipage}
	\end{figure*}

	If the rQSOs are more subject to dust extinction, then we would expect the optical AGN emission to be relatively weaker when compared to the $L_{\rm 6 \mu m}$ than the bQSOs and cQSOs.
	In Figure~\ref{fig:L6mu vs Lbol fraction} this is demonstrated by a comparison between the inferred bolometric luminosities from \citet[][]{shen2011} ($L_{\rm bol,Shen}$) to those derived from the $6\,\mu$m luminosity: $L_{\rm bol,6\mu m} = BC_{\rm 6\mu m}~\times~L_{\rm 6\mu m}$, where we adopted  $BC_{\rm 6\mu m} = 8$ from \citet{stanley2015}. 
	The continuum luminosities at rest-frame 5100\,\AA{} (left), 3000\,\AA{}  (middle) and 1350\,\AA{} (right) provided in \citet{shen2011} were used as the optical bolometric luminosity. 
	The median luminosity ratios (vertical dash-dot lines) computed from the 5100\,\AA{} luminosity are relatively consistent for the bQSOs, cQSOs and rQSOs.
	Contrariwise, as we approach shorter wavelengths, going from 3000\,\AA{} to 1350\,\AA{} luminosities, the median luminosity ratios of the red quasars decrease with respect to that of the blue quasars. 
	This is the signature that we would expect for dust reddening as the shorter wavelength emission will be more suppressed for a fixed amount of obscuration than longer wavelength emission.
	Conversely to the underestimated luminosities, a few red quasars have notable overestimated optical bolometric luminosities (lower tail) which may be due to host contamination in lower luminosity quasars at low redshifts.  
	At $z < 0.5$ the $\Delta (g^* - i^*)$ values of the blue and red quasars are seen to deviate from the overall population (see Figure~\ref{fig:delta_gi_fig}); the blue quasars have bluer colours than the median quasar at the specific redshift, whereas the red quasars have redder colours. A possible explanation for the change in the apparent shape of the distribution is likely host galaxy contamination at lower redshifts.
	%-----------------------------------------------------------------------------------------------------------------------
	%-----------------------------------------------------------------------------------------------------------------------
	
	\subsubsection{Defining the luminosity-redshift matched colour-selected quasar samples }
	\label{subsubsec:matched L6um-z}
	The different optical and MIR luminosity distributions between the bQSOs, cQSOs and rQSOs shown in \cref{subsubsec: WISE properties} could skew any differences in the intrinsic properties between red and blue quasars. We have therefore adopted a luminosity and redshift matching approach to refine our quasar samples and exclude any biases in our results introduced by luminosity differences.
	
	We luminosity matched the bQSOs, cQSOs and rQSOs in rest-frame $6\,\mu$m luminosity and redshift using tolerances of 0.2~dex and 0.05, respectively, with the following procedure (see Figure~\ref{fig:flowchart} and Table~\ref{tab:source_stats}): (1) we used the the 2-d Cartesian Anisotropic algorithm from the `Triple Match' function in {\sc topcat} \citep{taylor2005,taylor2006} keeping the best symmetric matches of the bQSOs, cQSOs and rQSOs in $L_{\rm 6\mu m}$ and redshift (bQSO$^{L_{\rm 6\mu m}}_{z}$, cQSO$^{L_{\rm 6\mu m}}_{z}$ and rQSO$^{L_{\rm 6\mu m}}_{z}$) and (2) we associated the quasars in these matched samples to FIRST within a 10$''$ radius (FbQSO$^{L_{\rm 6\mu m}}_{z}$, FcQSO$^{L_{\rm 6\mu m}}_{z}$ and FrQSO$^{L_{\rm 6\mu m}}_{z}$), as described in \cref{subsubsec:first data}. Overall, within the $L_{\rm 6\mu m}-{z}$ matched colour-selected sample there are 1967 quasars in each of the bQSO, cQSO, and rQSO sub samples, of which 129, 113, and 323 are radio detected, respectively.\\

	%==============================================================================================================
	%==============================================================================================================
	\section{Results}
	\label{sec:results}
	In this section we use our two defined samples (the full colour-selected sample and the $L_{\rm 6\mu m}-{z}$ matched colour-selected sample) to search for fundamental differences between the radio properties of red and blue quasars. We explore the radio-detection rates in \cref{subsec:radio-detection fraction result}, the radio morphologies in \cref{subsec:radio morphologies}, and the radio luminosities ($L_{\rm 1.4GHz}$ and $L_{\rm 1.4GHz}/L_{\rm 6 \mu m}$) in \cref{subsec:radio luminosities}.
	%==============================================================================================================
	\subsection{FIRST detection rates of red versus blue quasars} 
	\label{subsec:radio-detection fraction result}
\begin{figure*}
		\centering
		\includegraphics[width=35pc]{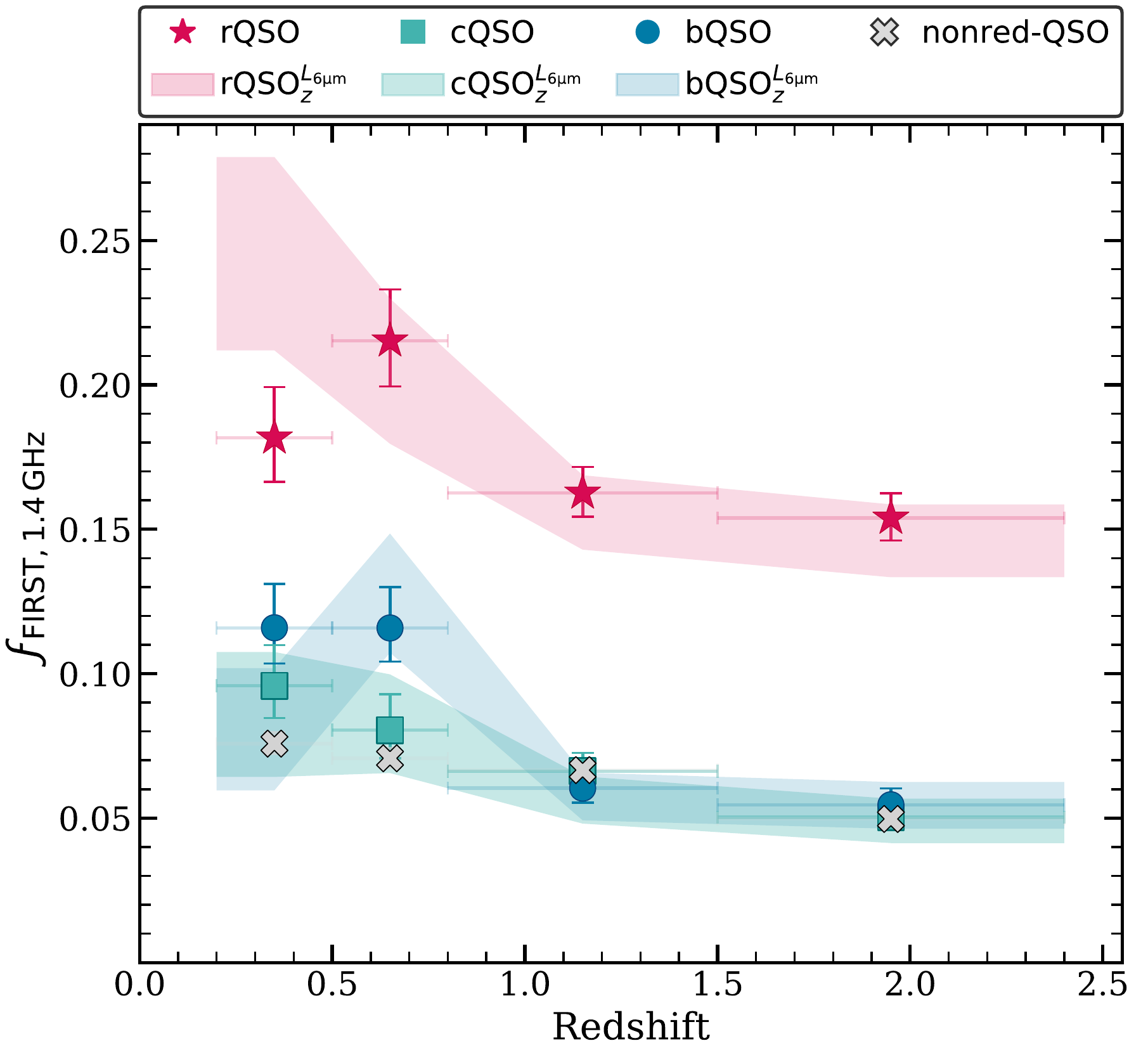}		
		\caption{The FIRST 1.4~GHz radio-detection fraction as a function of redshift for the blue (bQSOs), control (cQSOs) and red (rQSOs) quasars using the colour scheme in Figure~\ref{fig:gi_z_distribution}; we also plot on the radio-detection fraction for non-red QSOs, which is defined as all quasars excluding the rQSOs. The data are taken from Table~\ref{tab:radio detection frac} and the errorbars correspond to the 1$\sigma$ binomial uncertainties. The horizontal bars denote the redshift ranges of the four $z$-bins. The rQSOs have a factor $\approx$~2--3 times larger radio-detection fraction in comparison to the bQSOs and cQSOs. The shaded regions represent the FIRST detection rates for the $L_{\rm 6\mu m}-z$ matched bQSOs, cQSOs and rQSOs.}
		\label{fig:radio_detection_frac}	
	\end{figure*}

	The FIRST radio-detection rate as a function of the median redshift in each of the four redshift bins for the colour-selected samples is presented in Figure~\ref{fig:radio_detection_frac}.
	We used the Bayesian binomial confidence interval technique discussed in \citet{cameron2011} to calculate 1$\sigma$ uncertainties on the radio-detection fractions for both the full and $L_{\rm 6\mu m}-z$ matched colour-selected samples. 
	In Table~\ref{tab:radio detection frac} we show the source statistics for the different samples.
	
	The bQSOs and cQSOs have similar radio-detection fractions of $\approx$\,5\% -- 10\% across all redshifts. 
	However, it is apparent that rQSOs always exhibit a significantly higher FIRST detection rate ($\approx$\,15\% -- 20\%) relative to non-red quasars; i.e.,\ the radio-detection fraction of red quasars is a factor of $\approx$\,2 -- 3 times larger than blue quasars.
	This enhancement is also apparent if we compute the radio-detection fraction for non-red quasars in our sample; i.e.,\ all quasars in the $g^* - i^*$ colour distribution excluding the reddest 10\%. The FIRST detection rates for these quasars are consistent with those of the bQSOs and cQSOs and, hence, they represent the typical quasar population of which only 5--10\% are radio bright. 
	Our results are consistent between the full and $L_{\rm 6\mu m}-z$ matched colour-selected samples.  
	%..............................................................................................................
	%..............................................................................................................
	%..............................................................................................................
	\bgroup
	\def\arraystretch{1.6}
	\begin{table}
		\begin{center}
			\small
			\caption{\label{tab:radio detection frac} The percentage of bQSOs, cQSOs, and rQSOs detected with FIRST. We also report the the detection fraction for non-red quasars; i.e.,\ excluding the 10\% of the $g^* - i^*$ distribution. Also given are the 1$\sigma$ uncertainties on the radio-detection fractions obtained from Bayesian statistics. See Table~\ref{tab:source_stats} for the source statistics of each sample in the respective redshift ranges and the sample descriptions. }
			\begin{tabular}[c]{ccccc}
				\hline
				\hline
				Redshift & \multicolumn{4}{c}{FIRST-detected percentage (\%)}  \\
				& bQSO & cQSO & rQSO & non-red QSO\\
				\hline
				\hline
				$0.2 < z < 0.5$ &  $11.6_{-1.2}^{+1.5}$ &  $9.6_{-1.1}^{+1.4} $ &  $18.2_{-1.5}^{+1.8}$ &   $7.1_{-0.3}^{+0.4}$  \\  
				$0.5 < z < 0.8$ &  $11.6_{-1.2}^{+1.4}$ &  $8.0_{-0.9}^{+1.3} $ &  $21.5_{-1.6}^{+1.8}$ &   $6.7_{-0.2}^{+0.2}$  \\  
				$0.8 < z < 1.5$ &  $6.0_{-0.5}^{+0.6} $ &  $6.6_{-0.5}^{+0.6} $ &  $16.3_{-0.8}^{+0.9}$ &   $5.0_{-0.2}^{+0.2}$  \\      
				$1.5 < z < 2.4$ &  $5.5_{-0.5 }^{+0.6}$ &  $5.0_{-0.5 }^{+0.5}$ &  $15.4_{-0.8}^{+0.8}$ &   $7.6_{-0.4}^{+0.4}$  \\
				\hline
				\hline
			\end{tabular}
		\end{center}
	\end{table}
	\egroup 
	%............................................................................................................
	
\begin{figure*}
		\centering
		\includegraphics[width=35pc]{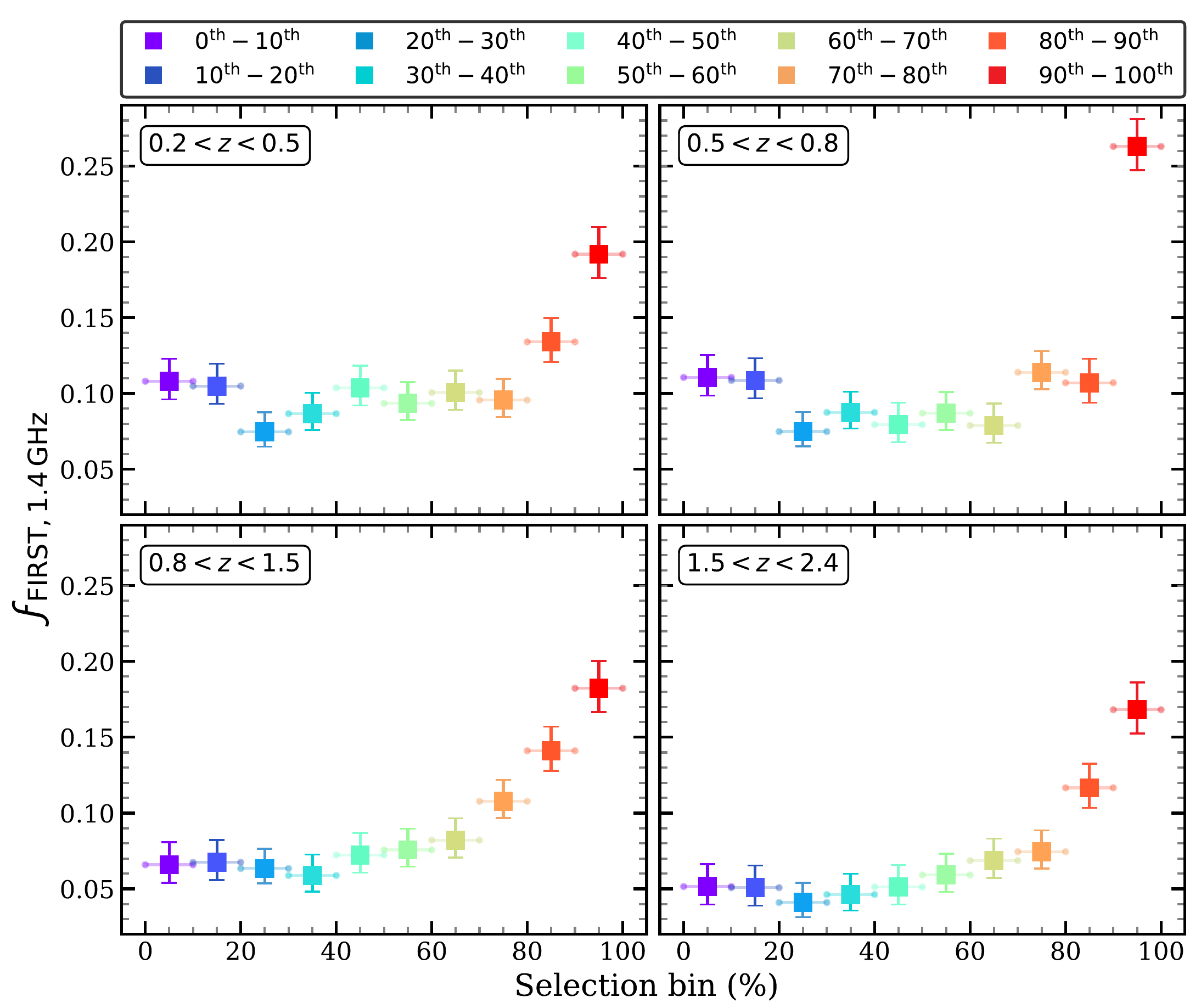}
		\caption{The FIRST 1.4~GHz radio-detection fraction as a function of the $g^* - i^*$ quantile split into 10\% selection bins in the four respective redshift bins. The top 10\% of the reddest SDSS quasars always show the largest radio-detection fraction. }
		\label{fig:selection_10percent_bins}
	\end{figure*}

	In our colour sample definitions we have taken the top 10\% to define rQSOs (see \cref{subsec:sample definition}). 
	In Figure~\ref{fig:selection_10percent_bins} we show how the radio-detection fraction changes from the bluest to reddest quasars: the top 10\% of the reddest SDSS quasars always show the largest radio-detection fraction; however, we note that a more refined colour-based selection of red quasars could result in even more significant differences from blue quasars.

	%==============================================================================================================
	\subsection{Radio morphologies of red versus blue quasars}
	\label{subsec:radio morphologies}
	Given the differences in the radio-detection fractions of red and blue quasars it is useful to determine whether there are associated differences in the radio morphologies; i.e.,\ whether red quasars favour a specific radio morphological class above another. 
	Radio-detected quasars often exhibit large scale radio jets and lobes which extend to tens of kpc, and in extreme cases can reach up to several Mpc \citep[e.g.,][]{muxlow1991}. 
	Their radio morphologies are diverse but can be broadly characterised into two groups: core and extended radio sources \citep[e.g.,][]{peterson1997,lu2007}. 
	Core radio sources are spatially unresolved at resolutions similar to FIRST and tend to have flat radio spectra ($\alpha < 0.5$). 
	In contrast, quasars with extended radio morphologies have radio spectra that are steep ($\alpha > 0.5$) and generally have the signature of two symmetric lobes (spatially resolved), although they can exhibit asymmetric lobes, jet-tail structures and other extended features.
	Extended radio sources are further differentiated into two main classes based on the surface brightness of the core to lobes, i.e.,\ Fanaroff-Riley (FR) type I and II \citep{fanaroffriley1974}, where the former are core-dominated systems and the latter are lobe-dominated systems \citep[e.g.,][]{fanaroffriley1974,bicknell1995}. % see Kimball+2011 
	
\begin{figure}
		\centering
		\includegraphics[width=20pc]{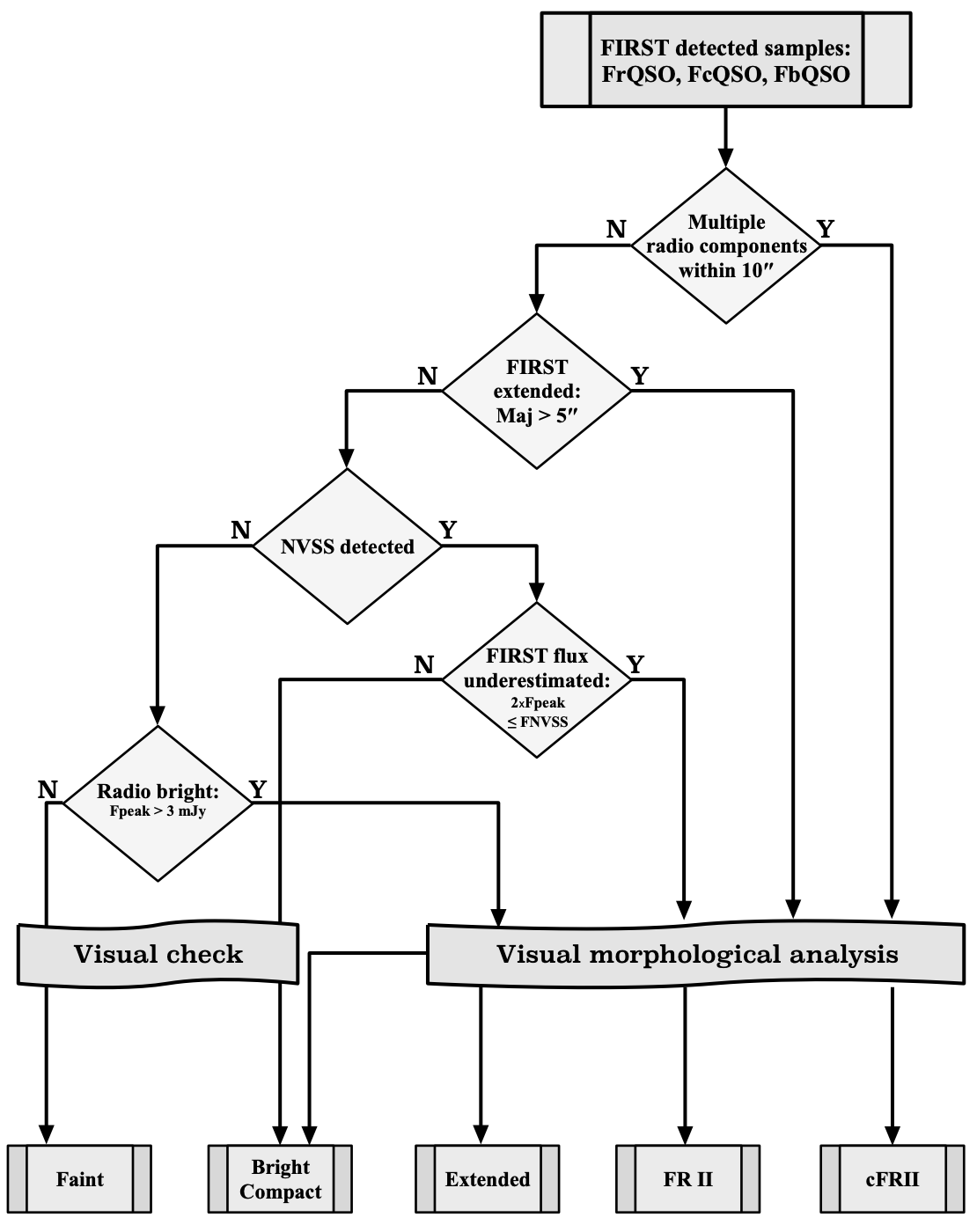}		
		\caption{Schematic flow diagram of our radio morphology classification for the FrQSOs, FcQSOs and FbQSOs. Four subcategories were selected to be visually assessed to accurately determine the morphological class. For completeness we also visually checked all of the other radio-detected quasars.}
		\label{fig:first morphology flowchart}	
	\end{figure}

	With the modest resolution of FIRST we can explore radio morphologies for sources that extend beyond the 5$''$ beam size of the survey (this corresponds to projected sizes in the range 17--41\,kpc at $z = 0.2-2.4$ for our assumed cosmology).
	However, given the diverse range of radio morphologies of quasars, some will be too extended to be picked up by FIRST. 
	Given these challenges we developed a comprehensive strategy to identify potentially extended radio sources, which we then visually classify using a simple approach that captures the morphological diversity of radio-detected quasars. 
	Our approach to identify and classify potentially extended sources is illustrated in Figure~\ref{fig:first morphology flowchart} and is described below.
	%..............................................................................................................
	%..............................................................................................................
	
	The FIRST survey decomposes radio sources into multiple gaussian components and reports size measurements for these components \citep{white1997}.
	Therefore the simplest approach in identifying extended sources is to search for cases where the gaussian sizes are larger than the FIRST beam size (5$''$) or with multiple radio counterparts within our cross-match radius of 10$''.$\footnote{Most of these FIRST-detected colour-selected quasars are associated with single radio counterparts within 10$''$ of the optical position.  
	Only 4\% of the FrQSOs have multiple radio components, whilst 8\% of the FbQSOs and 10\% of the FcQSOs have multiple radio components.}  
	This approach will identify the vast majority of extended sources but will miss the most extended systems that lack bright radio cores; i.e.,\ sources with very large lobe structures that lie beyond our 10$''$ cross-match radius and remain undetected by FIRST due to a faint radio core. To overcome this limitation we employed the 1.4~GHz NVSS survey which overlaps with the FIRST and SDSS footprint, but has a beam size of $45''$, significantly larger than FIRST. 
	This allows for the inclusion of large-scale radio structures beyond 10$''$. 
	%only two radio-detected sources in our sample have even larger lobe structures extending beyond 45$''$ (which corresponds to a physical size of $\approx$\,400~kpc at $z = 1.5$). 
	Using the methodology given in \citet{lu2007}, we estimate that only $\approx$\,0.2\% will have spurious matches or unassociated sources for $r$\,=\,10$''$ (this increases to $\approx$\,4\% for $r$\,=\,45$''$).
	We note that since NVSS is 2.5 times less sensitive than FIRST, faint diffuse extended sources will be missed in this analysis.
	%It should also be noted that many FIRST sources will not have NVSS counterparts because FIRST is a factor of 2.5 more sensitive than NVSS \citep{kimball2008}.
	
\begin{figure*}
			\centering
			\includegraphics[width=40pc]{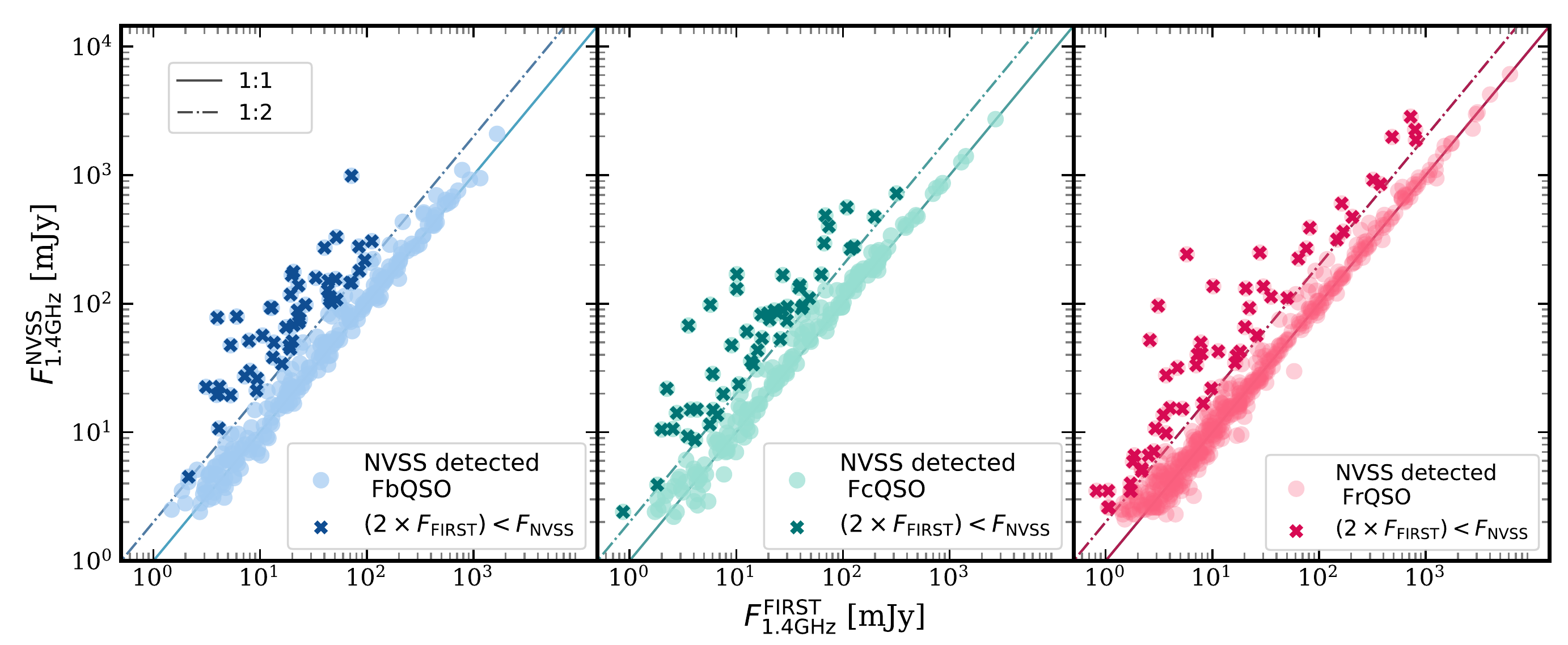}
			\caption{ NVSS flux vs. FIRST flux at 1.4\,GHz for the NVSS detected FbQSO, FcQSO and FrQSO indicated with faded blue, green and red circles. Sources with underestimated FIRST fluxes are indicated with dark blue, green and red crosses. The solid lines indicate a 1:1 relation and the dash-dot lines a 1:2 relation.}
			\label{fig:first summed fluxes}	
		\end{figure*}

	Figure~\ref{fig:first summed fluxes} shows the NVSS versus FIRST fluxes for the FbQSOs, FcQSOs and FrQSOs.
	The majority of the sources follow a 1:1 trend.   
	In this figure sources indicated with crosses have NVSS fluxes that are at least twice their FIRST fluxes: 12\%, 10\% and 7\% of the FbQSOs, FcQSOs and FrQSOs are highlighted. For these sources FIRST may have underestimated their extended emission and they were visually assessed (see Figure~\ref{fig:first morphology flowchart}).
	While radio variability may also explain these single component outliers \citep[e.g.,][]{heeschen1987,kraus2003,barvainis2005,czerny2008}, we nevertheless include them as potential extended sources. 
		%..............................................................................................................
		%..............................................................................................................
	
	All potentially extended sources highlighted by the process illustrated in Figure~\ref{fig:first morphology flowchart} were visually classified by three people (LK, DMA, DJR) using cutout images at 1.4~GHz extracted from the \href{https://third.ucllnl.org/cgi-bin/firstcutout}{VLA FIRST server}.  
	Overall we inspected 396 FIRST cutouts of 2$'$~$\times$~2$'$ fields centred on the S10 QSO positions. 
	This field size is sufficient since all but 7 of our radio-detected quasars have smaller angular sizes than 2$'$ \citep[see e.g.,][]{devries2006}, which corresponds to projected sizes of $<$~0.4--1.0\,Mpc at $z = 0.2-2.4$.
	%It should be noted that the observed radio morphologies can be redshift-dependent since some extended sources may be unresolved at higher redshifts \citep{kimball2011}.
	In our visual assessment we employed a simplified classification scheme that captures the main morphological classes of radio-detected quasars and minimises redshift- and resolution-dependent effects: (1) faint, (2) compact, (3) extended, (4) FR\,II-like and (5) compact FR\,II; see Table~\ref{tab:morphology classes} for a more detailed description. Of all the FIRST-detected quasars in our study, 27\% required visual assessment according to our approach with the remaining 73\% of radio-detected quasars categorized either as compact (non-extended; single components with fitted major axes less than 5$''$) or faint ($F_{\rm peak} < 3$\,mJy) radio sources. Although the compact and faint radio sources are not expected to be extended, for completeness we performed a visual check of their morphologies: we found 5\% of the faint and compact samples to have either extended or FR~II-like morphologies and we reclassified these systems.
	%..............................................................................................................
	\bgroup
	\def\arraystretch{1.3}
	\begin{table*}
		\begin{center}
			\footnotesize
			\caption{\label{tab:morphology classes} The radio morphology classes used to classify the FIRST images for the radio-detected quasars.   }
			\begin{tabular}[c]{llcc}
				\hline
				\hline
				Classification & Description \\
				\hline
				\hline
				\vspace{0.2cm}
				Faint & \makecell[l]{Sources detected close to the FIRST detection threshold with peak fluxes of $F_{\rm peak} < 3$\,mJy.} \\
				\vspace{0.2cm}
				Compact & \makecell[l]{Sources that are point-like with non-extended emission (fitted major axes $<$ 5$''$). } \\
				\vspace{0.2cm}
				Extended & \makecell[l]{Single radio sources that are extended. \\ This class includes FR\,I-like systems where both lobes are fainter than the core; $F_{\rm lobe} < F_{\rm peak}$.} \\
				\vspace{0.2cm}
				FR\,II-like & \makecell[l]{Double lobed systems with approximately the same brightness and offset from QSO position.  \\ At least one lobe should be brighter than than QSO core ($F_{\rm lobe} > F_{\rm peak}$).} \\
				\vspace{0.2cm}
				Compact FR\,II & \makecell[l]{FR\,II-like systems on small scales, i.e within the 10$''$ radius circle (which corresponds to a \\projected size of $\sim$\,85\,kpc at $z = 1.5$).} \\
				\hline
				\hline
			\end{tabular}
		\end{center}
	\end{table*}
	\egroup
	%..............................................................................................................
	
\begin{figure*}
		\centering
		\includegraphics[width=35pc]{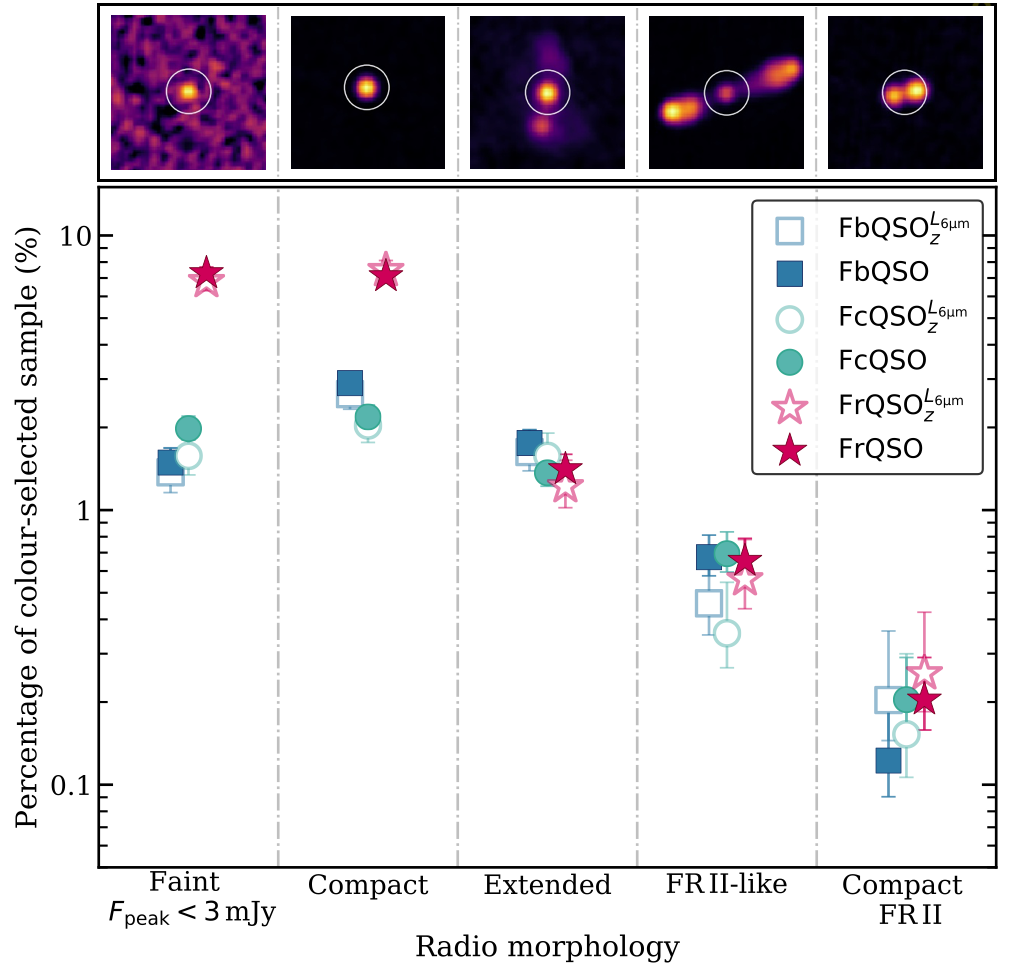}
		\caption{The percentage of both the full colour-selected (filled markers) and the $L_{6\mu m}-z$ matched (open markers) samples (using the colour scheme in Figure~\ref{fig:gi_z_distribution}) with different radio morphologies. The fractions are reported in five categories: faint sources detected near the sensitivity limit ($F_{\rm peak} < 3$\,mJy), bright compact radio sources, bright extended radio sources, FR\,II-like systems and compact FR\,IIs (small scale lobe-systems); see Table~\ref{tab:morphology classes} for more details. The errorbars correspond to the 1$\sigma$ binomial uncertainties and the vertical dash lines separate the different categories. Example 2$'$~$\times$~2$'$ FIRST images of each morphological class are illustrated in the top panel. The white circle represents our cross-matching radius of 10$''$. Extended radio emission is found among a similar fraction of all quasars, but red quasars show a surfeit of compact and faint systems.
		}
		\label{fig:first morphology summaries}	
	\end{figure*}

	Figure~\ref{fig:first morphology summaries} summarizes the results of our morphological analysis by showing the percentages of the bQSOs, cQSOs and rQSOs in each of the morphology classes; see also Table~\ref{tab:radio morph stats}. We calculated the 1$\sigma$ binomial uncertainties on our percentages for both the full and $L_{\rm 6\mu m}-z$ matched colour-selected samples following \citet{cameron2011}.
	This figure shows that rQSOs have a different radio morphological mix to blue quasars. There are no significant differences in the fraction of rQSOs and blue quasars with extended radio morphologies, such as the classical FRII-type systems. However, a factor 2--6 more rQSOs have either compact radio emission or are radio faint, in comparison to blue quasars, a result that is consistent between both the full and $L_{\rm 6\mu m}-z$ matched colour-selected samples. It is these compact and faint radio sources that are responsible for the increase in the overall radio-detection fraction between rQSOs and blue quasars in Figure~\ref{fig:radio_detection_frac}.
	%..............................................................................................................
	%..............................................................................................................
	%..............................................................................................................
	\bgroup
	\def\arraystretch{1.3}
	\begin{table}
		\begin{center}
			\small
			\caption{\label{tab:radio morph stats} The number and percentage of quasars in each of the morphology classes described in Table~\ref{tab:morphology classes}; see Figure~\ref{fig:first morphology summaries}. }
			\begin{tabular}[c]{llllcccc}
				\hline
				\hline
				Classification & bQSO & cQSO & rQSO  \\
				\hline
				\hline
				\multicolumn{4}{c}{The number of quasars in each class:} \\
				\hline
				Faint           & 73    & 97   & 357 \\      
				Compact         & 142   & 107   & 348 \\      
				Extended        & 86    & 67    & 69  \\      
				FR\,II-like     & 33    & 34    & 32  \\      
				Compact FR\,II  & 6     & 10     & 10   \\     
				\\
				\hline
				\multicolumn{4}{c}{The percentage in each class (see Figure~\ref{fig:first morphology summaries}):} \\
				& (\%) & (\%) & (\%) \\
				\hline
				Faint &          $1.49 _{- 0.153 }^{+ 0.193 } $ & $1.98 _{- 0.179 }^{+ 0.218 } $ & $7.287 _{- 0.352 }^{+ 0.387 }$ \\
                Compact &        $2.898 _{- 0.22 }^{+ 0.258 } $ & $2.184 _{- 0.189 }^{+ 0.228 }$ & $7.103 _{- 0.348 }^{+ 0.383 }$ \\
                Extended &       $1.755 _{- 0.168 }^{+ 0.207 }$ & $1.368 _{- 0.146 }^{+ 0.186 }$ & $1.408 _{- 0.148 }^{+ 0.188 }$ \\
                FR\,II-like &    $0.673 _{- 0.097 }^{+ 0.137 }$ & $0.694 _{- 0.099 }^{+ 0.139 }$ & $0.653 _{- 0.096 }^{+ 0.136 }$ \\ 
                Compact FR\,II & $0.122 _{- 0.032 }^{+ 0.073 }$ & $0.204 _{- 0.046 }^{+ 0.086 }$ & $0.204 _{- 0.046 }^{+ 0.086 }$ \\      
				\\
				\hline
				\hline
			\end{tabular}
		\end{center}
	\end{table}
	\egroup
	%..............................................................................................................

	%==============================================================================================================
	\subsection{Radio luminosities of red versus blue quasars}
	\label{subsec:radio luminosities}
	The excess of faint radio detections in red quasars compared to blue quasars suggests that a larger fraction of FrQSOs have low radio luminosities. 
	In addition to the radio morphologies we therefore also explored the radio luminosities ($L_{\rm 1.4GHz}$) and ``radio-loudness" (here defined as $L_{\rm 1.4GHz}/L_{\rm 6 \mu m}$) of the FIRST-detected quasars. 
	For consistency, in these analyses we used the total FIRST fluxes of all radio-detected components within 10$''$ to calculate the 1.4\,GHz luminosities, even if their NVSS fluxes differed from FIRST (see \cref{subsubsec:first data}).
	
\begin{figure*}
		\centering
		\includegraphics[width=40pc]{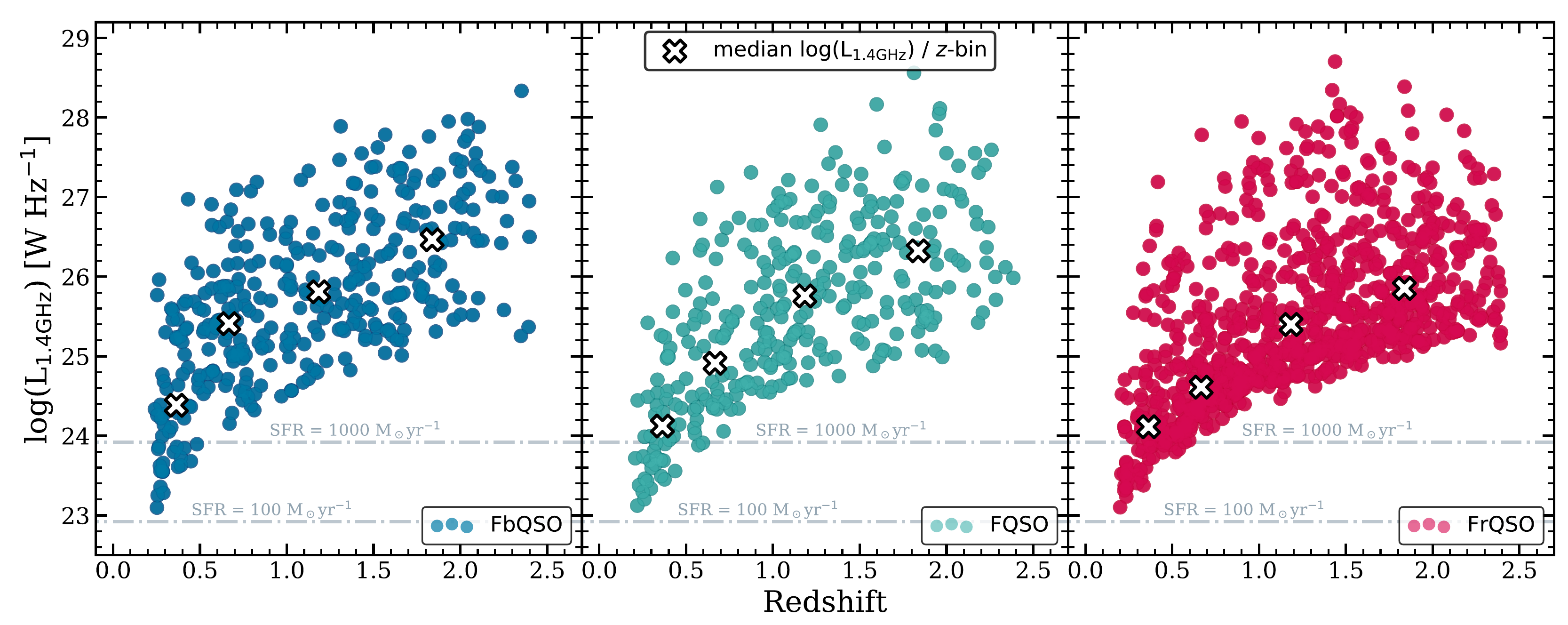}		
		\caption{Radio luminosity at 1.4\,GHz versus redshift for the radio-detected FbQSO (left), FcQSO (middle) and FrQSO (right) quasars. The FIRST fluxes of quasars with multiple radio counterparts within 10$''$ have been added to compute the total luminosity. The median $\log(L_{\rm 1.4GHz})$ in each of the 4 redshift bins are indicated with crosses. The dash-dot lines indicate radio luminosities corresponding to star formation rates (SFRs) of 100 and 1000 M$_\odot$\,yr$^{-1}$ \citep[see][for details on the SFR calculations]{hopkins2001}.}
		\label{fig:luminosity vs z}	
	\end{figure*}

	In Figure~\ref{fig:luminosity vs z} we plot $L_{\rm 1.4GHz}$ vs. redshift for the FbQSOs, FcQSOs and FrQSOs.
	The radio luminosity distributions of all three samples changes with redshift due to the flux limit of the FIRST survey. 
	Beyond $z > 0.5$, $L_{\rm 1.4GHz}$ is higher than that of the strongest star-forming galaxies, indicating that the radio emission must be dominated by the AGN in almost all cases \citep[e.g.,\ Arp~220 is the most powerful nearby starburst with $\log(L_{\rm 1.4GHz}/{\rm W~Hz^{-1}})$ = 23.4;][]{condon2013}.
	Noticeably, as indicated by our morphology analysis, there is a strong concentration of FrQSOs close to the FIRST detection limit. 
	%..............................................................................................................
	%..............................................................................................................
	
	An alternative way to investigate the radio power is to explore the relative ratio of the quasar in the radio band to the overall accretion power. This is effectively a measure of the ``radio loudness", which we define here as $R$.
	%However, rather than defining $R$ as the radio-to-optical flux ratio, as used in many other studies \citep[e.g.,][]{kellerman1989,urry1995,ivezic2002,glikman2004,zakamska2014,mehdipour2019}, in our analyses here we use the rest-frame 1.4~GHz--6~$\mu$m luminosity ratio, which will be less effected by dust extinction and redshift effects. 
	Many studies use the 5\,GHz-to-2500\,\AA{} flux ratio to define radio-loud quasars as objects with $R > 10$ \citep[e.g.,][]{kellerman1989,stocke1992, urry1995,ivezic2002,glikman2004,zakamska2014,mehdipour2019}. The transition between the radio-quiet and radio-loud regime is not definite and often quasars with $3 < R < 100$ are considered to be radio-intermediate. % see glikman+2004
	Using the $R$-values provided in \citet{shen2011} and the standard definition of radio loudness, we find that 81\%, 78\% and 83\% of the FbQSOs, FcQSOs and FrQSOs are radio-loud quasars with $R>10$.
	
\begin{figure*}
		\centering
		\includegraphics[width=35pc]{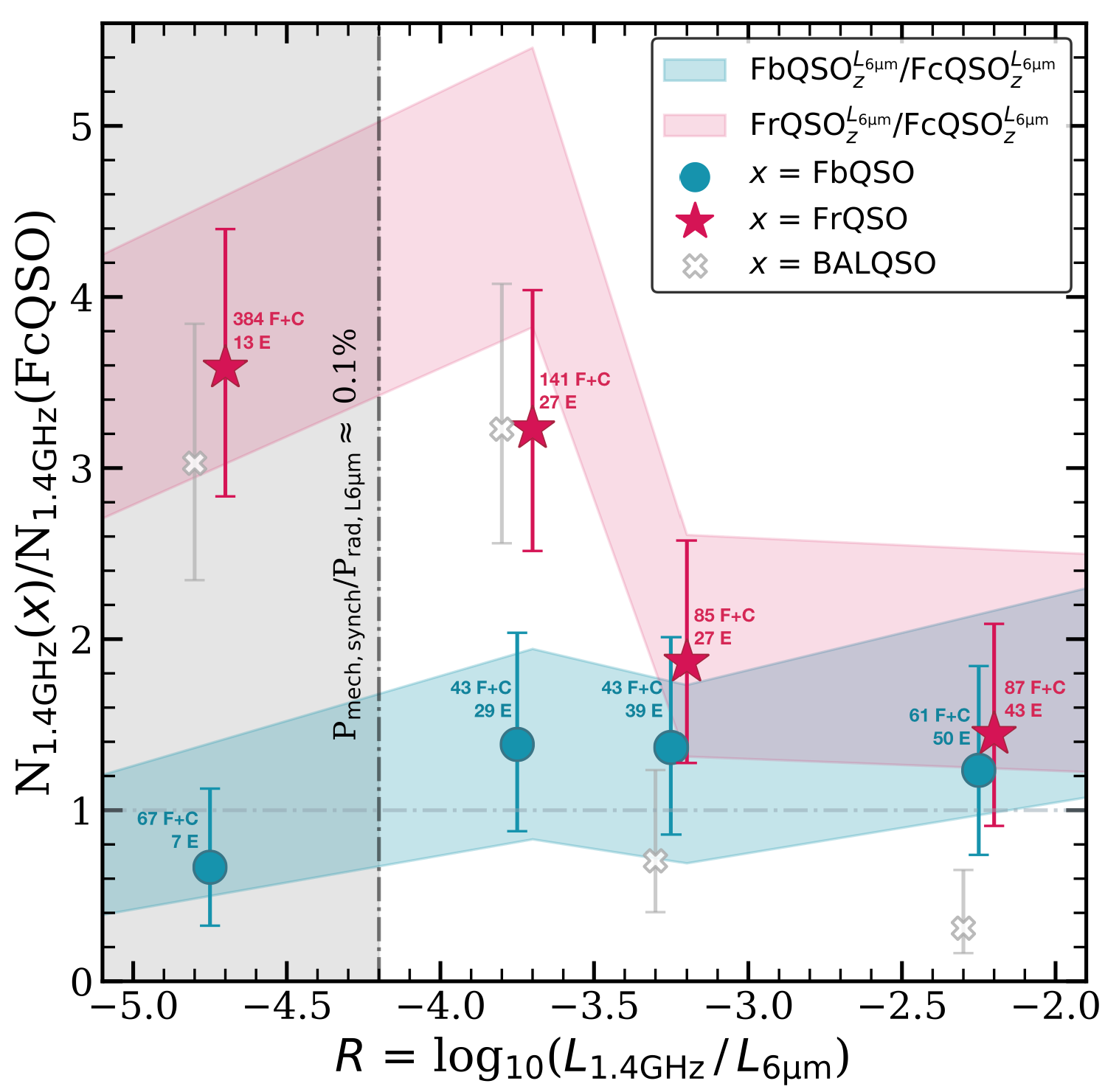}
		
		\caption{The FIRST 1.4~GHz radio-detection fraction in bins of radio loudness, computed using the 1.4\,GHz and 6\,$\mu$m luminosities, of the full (coloured markers) and $L_{\rm 6\mu m}-z$ matched (coloured shaded regions) colour samples. The relative fraction of FIRST-detected C\,{\sc iv} BALQSOs compared to FcQSOs at $1.5 < z < 2.4$ are also plotted as a function of $R$ (grey crosses; see \cref{subsec: empirical constraints evolution model}). The blue, red and grey markers are offset for illustration purposes. The grey shaded region indicates the radio-quiet regime and the dashed-dot line show the transition between radio-quiet and radio-loud quasars, quantified in terms of the mechanical-to-radiative power (see footnote~4 and \cref{subsec:radio luminosities}). The number of sources in each $R$-bin is subdivided into faint and compact morphologies (F+C) and extended (E; including extended, FR~IIs and compact FR~IIs) morphologies. The enhanced radio-detection fraction for the rQSOs is predominantly from systems around the radio quiet-radio loud threshold with faint or compact radio morphologies.}
		\label{fig:radio-loudness}
	\end{figure*}

	To reliably constrain the radio loudness in our analyses here we use the rest-frame 1.4\,GHz\,--\,6\,$\mu$m luminosity ratio ($L_{\rm 1.4GHz}/L_{\rm 6 \mu m}$), which will be less affected by dust extinction and redshift effects. 
	In Figure~\ref{fig:radio-loudness} we plot the relative fraction of FrQSOs and FbQSOs compared to FcQSOs as a function of $R$ for both the full (filled markers) and $L_{\rm 6\mu m}$--z matched (shaded regions) colour-selected samples.  
	We estimate the transition between radio-quiet and radio-loud quasars by requiring that 90\% of the \citet{shen2011} identified radio-loud FcQSOs satisfies our radio loudness definition: $\log_{10}(L_{\rm 1.4GHz}/L_{\rm 6 \mu m}) \approx -4.2$ (dot-dashed line); we used the FcQSOs since they will have typical optical quasar colours unaffected by dust reddening. 
	This figure shows a factor $\approx$~3--4 enhancement in the fraction of FrQSOs with respect to both FbQSOs and FcQSOs towards lower values of $R$, indicating that the enhanced radio-detection fraction for the FrQSOs is from systems around the radio-loud--radio-quiet threshold; we quantified this threshold ($\log_{10}(L_{\rm 1.4GHz}/L_{\rm 6 \mu m}) \approx -4.2$) in terms of the mechanical-to-radiative power to be $P_{\rm mech,sync}/P_{\rm rad,L6\mu m}\approx0.001$.\footnote{On the basis of the methodology given in \citet{willot1999}, we calculate the mechanical power from the jet and find that for our assumed radio quiet-radio loud threshold the mechanical-to-radiative power ($P_{\rm mech,sync}/P_{\rm rad,L6\mu m}$) corresponds to $\sim$\,0.1\%. In this calculation we assume a normalization factor $\digamma_{\rm W}$ = 5 \citep[see][]{daly2012}; the normalization factor ranges between 1\,--\,20, and combines several factors such as the lobe filling factor and the amount of energy from non-radiating particles.} As may be expected, given our earlier radio-morphology results, the enhancement in the fraction of rQSOs at low $R$ values comes from systems with either faint or compact radio morphologies; the number of sources in each bin are indicated in Figure~\ref{fig:radio-loudness} and tabulated in Table~\ref{tab:radio loudness source stats}. No significant differences are found between red and blue quasars within the classical extended radio-loud systems.
	%\citet{glikman2012} define the radio loudness parameter using the 1.4\,GHz and $g$-band fluxes and show that red quasars have similar radio distributions to blue quasars after correcting for dust extinction. Furthermore, in their sample a large number of red and blue quasars are radio-quiet.
	%..............................................................................................................
	\bgroup
	\def\arraystretch{1.3}
	\begin{table}
		\begin{center}
			\small
			\caption{\label{tab:radio loudness source stats} The number of FbQSOs, FcQSOs and FrQSOs in the bins of $\log_{10}(L_{\rm 1.4GHz}/L_{\rm 6 \mu m})$ plotted in Figure~\ref{fig:radio-loudness}. Each bin is subdivided into our radio morphology classes; faint, compact and extended sources (extended, FR~II and compact FR~II).   }
			\begin{tabular}[c]{lllllccccc}
				\hline
				\hline
				$\log_{10}(L_{\rm 1.4GHz}/L_{\rm 6 \mu m})$ & Sample & Faint & Compact & Extended \\
				\hline
				\hline
				-4.75 & FbQSO & 53 & 14 & 7 \\
				& FcQSO & 84 & 16 & 11 \\
				& FrQSO & 321 & 63 & 13 \\
				\hline
				-3.75 & FbQSO & 17 & 26 & 29 \\
				& FcQSO & 11 & 25 & 16 \\
				& FrQSO & 33 & 108 & 27 \\
				\hline
				-3.25 & FbQSO & 2 & 41 & 39 \\
				& FcQSO & 0 & 26 & 34 \\
				& FrQSO & 0 & 85 & 27 \\
				\hline
				-2.25 & FbQSO & 0 & 61 & 50 \\
				& FcQSO & 0 & 40 & 50 \\
				& FrQSO & 0 & 87 & 43 \\
				\hline
				\hline
			\end{tabular}
		\end{center}
	\end{table}
	\egroup
	%..............................................................................................................
	%..............................................................................................................
	%..............................................................................................................
	
	\section{Discussion}
	\label{sec:discussion}
	We have analysed the SDSS DR7 Quasar Catalogue from S10 to look for fundamental differences in the radio properties between red and blue quasars.
	By carefully selecting rQSOs, cQSOs and bQSOs from the top, middle and bottom 10\% of the observed $g^* - i^*$ colour distributions we have generated uniformly selected samples unbiased in their radio properties. Overall, we have found that rQSOs have a FIRST radio-detection fraction of $\approx 15 - 20$\%, which is a factor of $\approx\,2-3$ times larger than that of blue quasars (cQSOs and bQSOs). Through a visual inspection of the FIRST images and an assessment of the radio luminosities (${L_{\rm 1.4~GHz}}$ and ${L_{\rm 1.4~GHz}}/{L_{\rm 6\mu m}}$) we find that the radio-detection excess for rQSOs is primarily due to compact and radio-faint quasars (those around the radio quiet-radio loud threshold). No significant differences are found between rQSOs, cQSOs, and bQSOs within the classical extended radio-loud systems. We find consistent results between our full and $L_{\rm 6\mu m}$--z matched colour-selected samples. Given that the radio luminosities of the quasars are at least an order of magnitude above those expected from star-formation (e.g.,\ $L_{\rm 1.4GHz} > 10^{25}$\,W\,Hz$^{-1}$ at $z > 1.5$), the differences in the radio properties will be driven by AGN-related processes (e.g.,\ radio core, jets, lobes, winds). 
	
	Many previous studies have explored the optical properties of radio-detected quasars, finding that they tend to have redder optical colours than radio-undetected quasars \citep[e.g.,][]{ivezic2002,white2003}.
	While other studies have explored the radio morphologies of radio-detected quasars and found that the quasars with unresolved radio emission tend to have redder optical colours than the quasars with extended radio emission \citep[e.g.,][]{lu2007,kimball2011}.
	However, fewer studies have explored the radio properties of carefully selected red and blue quasars.
	\citet[][]{tsai2017} distinguished between red and blue quasars based on their spectral colours (flux ratio of the rest frame 4000\AA{} to 3000\AA{} continuum emission) and, similarly to our work, found a larger number of red quasars to be detected with FIRST; however, their analysis was restricted to low redshifts ($0.3 < z < 1.2$). 
    Similarly, \citet{richards2003} reported that dust-reddened SDSS quasars have average FIRST detection-fractions $\approx$\,2\,$\times$ larger than intrinsically blue quasars while \citet{white2007} stacked the FIRST data of SDSS quasars as a function of $g^*-r^*$ colour and found an increase in the radio-flux density towards redder optical colours. We also note that \citet{georgakakis2009} tentatively found that a higher fraction of 2MASS NIR selected quasars are associated with radio emission than optically selected SDSS quasars.  \\
	%Furthermore, \citet{ross2015} examined the quasars with the reddest optical-to-MIR ($WISE$) colours in the SDSS quasar survey and found that 56\% of the extremely red quasars were detected in FIRST, again, pointing to some connection between reddening and radio that echoes our findings. 

	However, our study is the first to systematically explore the optical colour dependence on the radio-detection fraction as a function of radio morphology, luminosity, and ``radio-loudness". 
	A strength of our approach is that we have carefully selected comparison samples (rQSO, cQSO, and bQSO) from the same quasar population, allowing us to rule out significant selection and luminosity effects in our analyses.
	
	Overall, given the connection between the rQSOs and radio emission it is natural to ask whether they are produced by the same physical process. The only process likely to contribute to both the optical and radio wavebands is synchrotron radiation, which is the dominant physical mechanism in the radio waveband. Qualitatively, it appears unlikely that synchrotron radiation can explain the enhanced radio emission in rQSOs since the significant difference between rQSOs and blue quasars occurs in comparatively radio-weak systems. Indeed, from a more quantitative analysis, where we scale a synchrotron-dominated SED to the radio flux, we predict that only $\approx$~6\% of the FbQSOs and FcQSOs and $\approx$~8\% of FrQSOs are likely to suffer from significant contamination of the optical emission from synchrotron processes; see \cref{sec:appendix} for more details. Our results are qualitatively similar to those of \citet{glikman2007} who undertook a similar analysis but for a brighter NIR selected red-quasar sample.
	%Similarly, using radio spectral indices, \citet{glikman2007} %found that synchrotron emission contributes to the $K$-band %fluxes in $\sim$\,20\% of their radio-bright red quasars and %that there is no significant difference in these values between %their red and blue quasars. Our results are qualitatively %consistent with \citet{glikman2007}, but given the different %sample definitions they find a larger fraction of quasars with %significant contributions.
	We therefore conclude that dust extinction is the most plausible explanation for the optical colours of the majority of red quasars, in agreement with previous works \citep[e.g.,][]{webster1995,gregg2002,richards2003,glikman2004,glikman2007,glikman2012,kim2018}, particularly at $z > 0.5$ where dilution from the host galaxy is likely to be weak (see \cref{subsubsec: WISE properties}).
	
	On the basis of our results how can we explain the connection between the red colours and different radio properties of the rQSOs when compared to blue quasars? Below we discuss our results within the context of the two competing models for red quasars, the orientation model (\cref{subsec: evidence agains the orientation model}) and the evolutionary model (\cref{subsec: empirical constraints evolution model}).
	
	%=======================================================================================================================
	
	\subsection{Evidence against a simple orientation-dependent model}
	\label{subsec: evidence agains the orientation model}
	In a simple orientation-dependent model we would not expect physical differences between red and blue quasars; in this model any observed differences would merely be a consequence of more dust along the line-of-sight.
	Since radio emission is not affected by dust, we would therefore expect no significant differences in the radio-detection fraction between different quasar sub populations, in stark contrast to our result (factor $\approx$~3 times more radio-detected rQSOs when compared to bQSOs and cQSOs, even when controlling for luminosity and redshift effects). In fact, since blue quasars would be more face on than the red quasars, on the basis of the orientation model, we would actually expect the opposite result (i.e.,\ a relatively larger fraction of radio-detected bQSOs and cQSOs than rQSOs).
	
	The differences in the radio morphologies and the radio luminosities of the rQSOs and blue quasars also argue against the orientation model. In the orientation model the rQSOs are more inclined than blue quasars and so the radio emission would be more extended (on average) while we find an excess of compact radio morphologies for the rQSOs, again the opposite to what we would expect. The larger fraction of rQSOs with low radio luminosities (i.e.,\ either ${L_{\rm 1.4~GHz}}$ or ${L_{\rm 1.4~GHz}}/{L_{\rm 6\mu m}}$) is also inconsistent with the orientation model which would predict no significant differences in the radio luminosities of the different quasar sub populations.
	
	\subsection{Empirical constraints for an evolutionary model}
	\label{subsec: empirical constraints evolution model}
	%\textcolor{debianred}{Beta suggests moving this to the end of previous subsection: Our results argue that the differences between the radio properties of red and blue quasars cannot be solely explained by orientation. Do you agree Dave? I'LL UPDATE NOW LIZELKE - I'M JUST ON THIS POINT, OKAY?} 
	Since our radio results cannot be explained (solely) by orientation, they must imply fundamental differences between red and blue quasars. These  differences could be driven by changes in the physical properties of the quasars themselves (e.g.,\ the BH accretion disc) or they could be related to the larger-scale ``environment"; we use the term ``environment" here to describe a broad range of potential physical phenomena over a wide range of size scales (e.g.,\ pc--Mpc scales). We briefly discuss the empirical constraints that our study can place on these scenarios below.
	
	Several studies have found that a small number of X-ray, optical, and NIR selected quasars are characterized by intrinsically red continua, potentially related to different accretion rates in the red quasars compared to blue quasars \citep{puchnarewicz1998,richards2003,young2008,ruiz2014,kim2018}.
	%Comparable results have also been found by \citet{richards2003} and \citet{kim2018} for a minority of the quasar population from optical- and/or NIR-selected samples.
	%and found that a reddened continuum for a minority of the quasars in their samples can be explained by intrinsic continuum differences.
	Since it has also been noted that radio emission is enhanced in low Eddington rate AGN \citep[e.g.][]{heckman2004,kauffmann2008}, it is therefore reasonable to consider whether the rQSOs have lower accretion rates in comparison to bQSOs and cQSOs due to their enhanced radio-detection fractions.
	We test this here by estimating the range in Eddington ratios between the $L_{\rm 6\mu m}-z$ matched colour samples using the bolometric luminosities (estimated from $L_{\rm 6\mu m}$; see \cref{subsubsec: WISE properties}) and the FWHM of the broad lines as a proxy of the BH masses (i.e.,\ based on virial BH masses).
	%The virial BH masses are calculated using the optical continuum luminosity and broad-line components of H$\beta$, Mg\,{\sc ii} or C\,{\sc iv} depending on the redshift. 
	%Therefore, any variations in the FWHM measurements of the broad lines between the different subsamples will result in different estimated BH masses. DMA: I don't think we need to say this as we ultimately constrain the Eddington ratios rather than BH mass and I'm worried that we will get in a real twist if we seem to estimate BH mass as well
	
\begin{figure*}
		\centering
		\includegraphics[width=40pc]{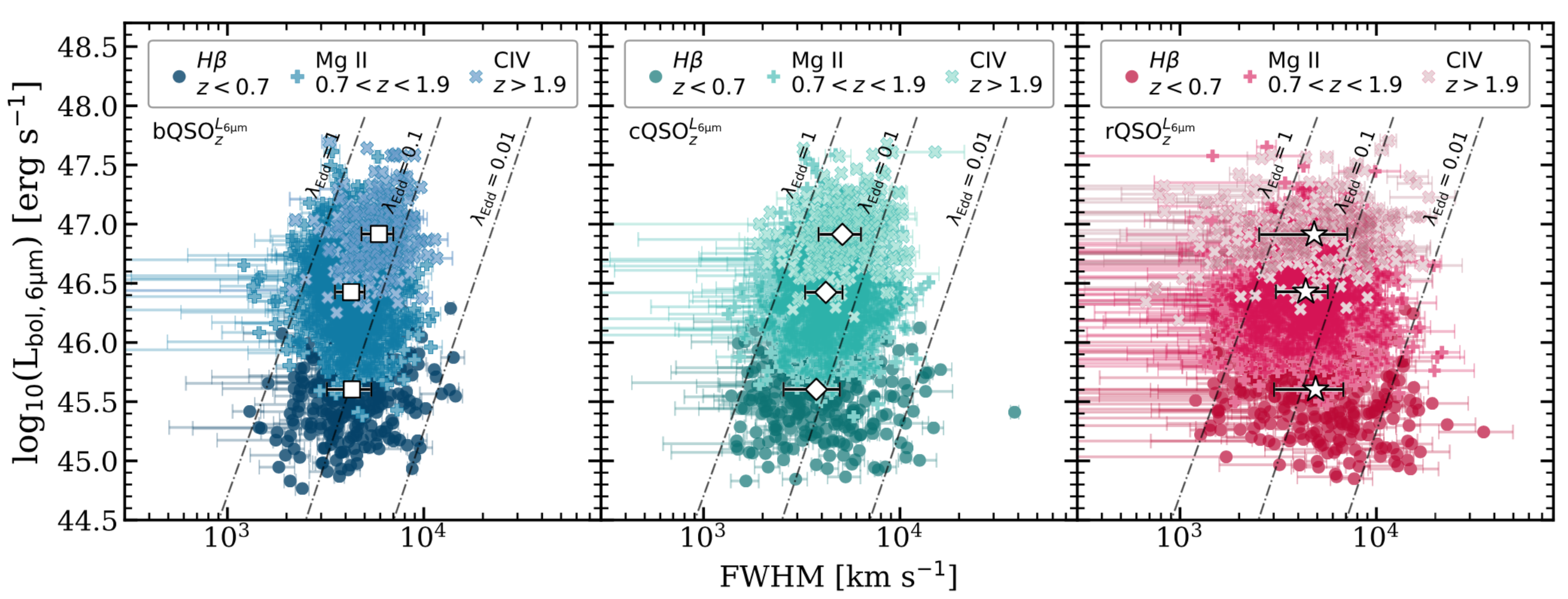}
		\caption{The bolometric luminosity estimated using the $L_{\rm 6\mu m}$ vs. FWHM of the broad-lines H$\beta$ (circles), Mg~{\sc ii} (crosses) and C~{\sc iv} (X's) of the b,c,rQSO$^{L_{\rm 6\mu m}}_{z}$ quasars. 
		The median values and median absolute deviations (MAD) are plotted for each emission line. Also plotted are Eddington ratio tracks (dash-dot grey lines) determined from the Eddington ratios provided \citet{shen2011} for the cQSOs (since they represent typical quasars). There are no strong differences in the FWHM measurements and implied Eddington ratios between the red quasars and blue quasars.}
		\label{fig:fwhm all matched}
	\end{figure*}

	In Figure~\ref{fig:fwhm all matched} we show the estimated bolometric luminosity ($L_{\rm bol,6\mu m}$) as a function of FWHM for the b,c,rQSO$^{L_{\rm 6\mu m}}_{z}$ quasars.
	The median values for the different broad lines (different $z$-bins) are overlaid and we calculated the median absolute deviation (MAD) as the uncertainty on the median.
	Eddington ratio tracks ranging from $0.01 < \lambda_{\rm Edd} < 1$ are also plotted, calibrated against values from \citet{shen2011} for the cQSOs.
	A large scatter in the FWHMs is observed, especially for the rQSOs which often have larger associated uncertainties due to lower SNR spectra. Nevertheless, from the median values it is clear that there are no statistically significant differences in the FWHM values between the b,c,rQSO$^{L_{\rm 6\mu m}}_{z}$ quasars. These results would appear to be in disagreement with \citet{richards2003} who found a trend of narrower Balmer lines towards redder optical colours; however, we note that this result is based on composite quasar spectra with higher SNR allowing them to see more subtle trends than in our analysis. 
	Using the bolometric and BH mass relation for the Eddington ratio, we estimated the median Eddington ratio for each bin of the bQSOs, cQSOs, and rQSOs: all are broadly consistent with an increase from low-to-high redshift of $\lambda_{\rm Edd} \approx $~0.1 -- 0.5. 
	Since the quasars are matched in luminosity we therefore argue that there are no strong differences in the {\it average} accretion rates between red and blue quasars. This result appears to be in disagreement with some previous studies which have found that radio-selected red quasars have higher accretion rates than blue quasars \citep[e.g.,][]{urrutia2012,kim2015_accretion}, on the basis of optical--NIR spectra of small quasar samples. However, our current analysis is limited by the large uncertainties in individual FWHM measurements from \citet{shen2011}. We aim to clarify the connection between red quasars and accretion properties in upcoming work using a matched sample of rQSOs and cQSOs with high-quality VLT-XSHOOTER optical--NIR spectroscopy. %\textcolor{debianred}{Does this look okay Lizelke? I've modified the comparison text to other studies.}
	
	%We also note that, whilst it is true that AGN in "radio mode" are found to have low Eddington ratios, \citet{urrutia2012} and \citet{kim2015_accretion} both found that radio-selected red quasars have exceedingly high accretion rates, and relatively low BH masses from detailed analyses of small samples of quasars. 
	%Our analysis is unfortunately limited due by the large uncertainty in the individual FWHM values. Upcoming work will focus on constraining the BH masses and consequently the Eddington ratios using NIR spectra.
	
	%However, our analyses for the rQSOs are limited due to the larger uncertainty in the individual FWHMs due to typically lower SNR spectra. Upcoming work with higher-quality VLT-XShooter optical-NIR spectra will aim to further explore the spectral properties of red quasars in comparison to blue quasars to better constrain the BH masses and accretion disc properties.
	%-----------------------------------------------------------------------------------------------------------------------
	%-----------------------------------------------------------------------------------------------------------------------
	
	While our analyses of the accretion properties do not show significant differences between the red and blue quasars {\it on average}, the enhancement in the fraction of rQSOs with compact or faint radio emission suggest fundamental differences between these sub populations. The radio compactness indicates that the source of the differences in the radio properties emerges on nuclear or galaxy scales ($<40$\,kpc). 
	Within the context of the evolutionary model, this could imply that the red quasars are in a younger transitional phase with small, but expanding, radio jets \citep[such as Compact Steep Spectrum (CSS) and gigahertz-peaked spectrum (GPS); see also e.g.,][]{odea1997, murgia1999, rossetti2006, randall2011, dallacasa2013, orienti2016}; indeed, \citet{georgakakis2012} has argued a similar interpretation for a small sample of FIRST-detected dust-reddened quasars.
	In this model red quasars represent an early obscured phase during which energetic winds drive away the obscuring dust and gas revealing an unobscured view of the accretion disc (i.e.,\ a blue quasar).
	As the dust cocoon expands it confines a young radio source which remains compact on galaxy scales.
	Strong shocks associated with these interactions could sustain their radio synchrotron emission \citep[e.g.,][]{hwang2018}, leading to the enhanced radio-detection fractions which we observe. 
	
	A prediction of the evolutionary model is that red quasars will host more powerful winds than blue quasars. Interestingly, \citet{mehdipour2019} recently showed an anti-correlation between the column density of ionised winds in quasars and the ``radio-loudness", whereby the systems with the weaker radio emission have stronger winds. The enhancement in the fraction of rQSOs with low $R = {L_{\rm 1.4~GHz}}/{L_{\rm 6\mu m}}$ values therefore indirectly suggests that they will host stronger winds than bQSOs. 
	On the basis of the current data, we are unable to test whether the rQSOs have stronger winds than the bQSOs. 
	However, we can use the sub population of broad absorption line quasars \citep[BALQSOs;][]{foltz1987,weymann1991}, which are known to host powerful winds, to see whether they are preferentially radio weak and to therefore provide an indirect test of this relationship.  
	In Figure~\ref{fig:radio-loudness} we plot the relative fraction of FIRST-detected BALQSOs compared to FcQSOs as a function of $R$; we identified BALQSOs as systems with C~{\sc iv} ($1.5 < z < 2.4$) broad absorption lines from our parent sample (see Figure~\ref{fig:flowchart}) using the data provided in \citet{shen2011} ({\sc bal\_flag} = 1 or {\sc bal\_flag} = 3). 
	We find an enhancement of BALQSOs at low values of $R$ and a deficiency at high values of $R$, results which are in general agreement with those obtained by \citet{morabito2019} for deeper low-frequency LOFAR radio data. The behaviour at high values of $R$ is strikingly similar to that found for the rQSOs, suggesting that rQSOs may also host powerful winds, in good agreement with the evolutionary model; see also \citet{urrutia2009} for evidence of an enhancement of BALQSOs in the red-quasar population.
	We will more directly test this hypothesis in a future VLT-XSHOOTER spectroscopic study and also determine whether the red quasars from our sample host more powerful winds than the blue quasars. 
	%If red quasars represent a brief evolutionary phase how does it link to BH spn...
	%If there is a playoff between wind and jet radiation due to different magnetic field configurations, one might expect it to be a result of black hole spin. \citet{chiaberge2015} for example found a relation between radio-loud AGN and merger fraction. However, this is inconsistent with our ``idea" of what is going on because we say the difference is in the radio-quiet phase. But perhaps this comes down to redshifts again and we study different objects at different epochs. Dave, I know this is too much detail, I'm just trying to lay it out in my mind. Link accretion dominated systems and winds to mergers -- but studies rather find a link between jet dominated systems and mergers.
	%Require HST data to study merger fraction in rQSOs and cQSOs. 
	%	However, we caution that we cannot confirm or refute such a model with the current data and more detailed observations are required to provide conclusive constraints.

	%================================================================================================================================
	\section{Conclusions}
	\label{sec:conclusion}
	We have taken a novel approach to search for fundamental differences between red and blue quasars at $0.2 < z < 2.4$ to allow us to test between the two main competing models: the unified orientation model and an evolutionary model.
	Our quasar selection is based on the SDSS survey and is uniformly selected and unbiased in the radio. We distinguished between red (rQSOs), control (cQSOs) and blue (bQSOs) quasars using carefully constructed $g^*-i^*$ colour distributions as a function of redshift. 
	Our rQSO selection is therefore sensitive to the redshift evolution of quasar SEDs. We found that the red colours of the rQSOs suggest that the majority require $A_{\rm V} \sim 0.1-0.5$\,mag of dust reddening to the median cQSO to produce the observed optical colours.
	%Similar results are found whether we use a colour-uniform, radio-uniform or luminosity-uniform sample (see \cref{subsec:radio-detection fraction} and \cref{subsec:matched L6um-z}).
	
	From a systematic comparison of the radio properties of the rQSOs, cQSOs and bQSOs, we have identified fundamental differences that cannot be attributed to just the orientation model. Our results are consistent between our full and $L_{\rm 6\mu m}-z$ matched colour-selected samples:
	\begin{itemize}
		\item{{\bf Enhanced radio emission from the AGN in rQSOs (see Figure~\ref{fig:radio_detection_frac}):} a larger fraction of rQSOs (by a factor of 2--3) are detected in the 1.4~GHz FIRST survey when compared to cQSOs and bQSOs. The average FIRST detection rate across $0.2 < z <2.4$ are 5\%--10\% for the cQSOs and bQSOs and 15\%--20\% for the rQSOs. The radio luminosities are at least an order of magnitude above those expected from SF (e.g.,\ $L_{\rm 1.4GHz} > 10^{25}$\,W\,Hz$^{-1}$ at $z > 1.5$), indicating that they are driven by AGN-related processes. See \cref{subsec:radio-detection fraction result}.
		%We found our result holds for both our full and $L_{\rm 6\mu m}-z$ matched colour-selected samples (\cref{subsec:radio-detection fraction result}).   
		}
		\item{{\bf rQSOs have differences in their arcsecond-scale radio morphologies (see Figure~\ref{fig:first morphology summaries}):} from a visual assessment of the FIRST cutouts we have found that the incidence of systems with extended and FR~II-like radio morphologies among the rQSOs is the same as among the cQSOs/bQSOs, but they have a much higher incidence of faint and compact radio counterparts (by a factor of 2--6). The radio-detection enhancement of the rQSOs therefore occurs in the compact and faint radio sources rather than the classical extended radio-loud systems. See \cref{subsec:radio morphologies}. 
			%The rQSOs also have different radio luminosity distributions to cQSOs/bQSOs with a larger fraction hosting radio faint radio sources (see \cref{subsec:radio luminosities}).
		}
		\item{{\bf rQSOs have lower radio--MIR luminosity ratios (${L_{\rm 1.4~GHz}}/{L_{\rm 6\mu m}}$; see Figure~\ref{fig:radio-loudness}):} we found a factor $\approx$~3--4 enhancement of rQSOs at low ${L_{\rm 1.4~GHz}}/{L_{\rm 6\mu m}}$ values, around the radio quiet-radio loud threshold, when compared to cQSOs and bQSOs. These differences are dominated by the compact and faint radio sources that are responsible for the enhanced radio-detection fraction of rQSOs. We see no significant differences in the classical extended radio-loud systems. See \cref{subsec:radio luminosities}.
		}
	\end{itemize}
	
	By linking the enhanced radio-detection rates and dust extinction of rQSOs we conclude that these sources are not blue quasars with additional dust along the line-of-sight due to a larger viewing angle as a simple orientation model would predict. By comparison we argue that the radio properties of the rQSOs are consistent with the evolutionary paradigm where rQSOs contain younger, more compact radio sources, possibly in a brief transitional phase where powerful winds are driving away the obscuring gas and dust.  
	
	In future work we will investigate radio spectral indices with multi-frequency radio data, which provide a more comprehensive understanding of the radio properties of the rQSOs.
	We will also further explore the origin of redness in the rQSOs from optical--NIR spectral analysis, constrain the star-formation rates from the host galaxy using {\it Herschel}--ALMA far-IR and mm imaging, and search for merger-driven signatures from high spatial resolution optical--NIR imaging.

	%==============================================================================================================
	%==============================================================================================================
	%%%%%%%%%%%%%%%%%%%%%%%%%%%%%%%%%%%%%%%%%%%%%%%%%%
	\section{Acknowledgements}
	We would like to acknowledge Elizabeth Wetherell for her assistance in the original discovery of the enhanced radio-detection fraction for red quasars. We also thank the anonymous referee for their useful comments which greatly improved the presentation and discussion of the results. We would also like to express our gratitude to the following people for their concise feedback and contributions: Manda Banerji, Alastair Edge, Richard McMahon, Andrea Merloni, Adam D. Myers, Gordon T. Richards, Nicholas P. Ross and Benny Trakhtenbrot. 
	
	We acknowledge the Faculty of Science Durham Doctoral Scholarship (LK), the Science and Technology Facilities Council (DMA, DJR, through grant code ST/P000541/1), a European Union COFUND/Durham Junior Research Fellowship (EL, through EU grant agreement no. 609412) and the financial support from the Swiss National Science Foundation (SF).
	
	Funding for the SDSS and SDSS-II has been provided by the Alfred P. Sloan Foundation, the Participating Institutions, the National Science Foundation, the U.S. Department of Energy, the National Aeronautics and Space Administration, the Japanese Monbukagakusho, the Max Planck Society, and the Higher Education Funding Council for England. The SDSS Web Site is http://www.sdss.org/.
	The SDSS is managed by the Astrophysical Research Consortium for the Participating Institutions. The Participating Institutions are the American Museum of Natural History, Astrophysical Institute Potsdam, University of Basel, University of Cambridge, Case Western Reserve University, University of Chicago, Drexel University, Fermilab, the Institute for Advanced Study, the Japan Participation Group, Johns Hopkins University, the Joint Institute for Nuclear Astrophysics, the Kavli Institute for Particle Astrophysics and Cosmology, the Korean Scientist Group, the Chinese Academy of Sciences (LAMOST), Los Alamos National Laboratory, the Max-Planck-Institute for Astronomy (MPIA), the Max-Planck-Institute for Astrophysics (MPA), New Mexico State University, Ohio State University, University of Pittsburgh, University of Portsmouth, Princeton University, the United States Naval Observatory, and the University of Washington.
	
	This publication makes use of data products from the Wide-field Infrared Survey Explorer, which is a joint project of the University of California, Los Angeles, and the Jet Propulsion Laboratory/California Institute of Technology, funded by the National Aeronautics and Space Administration.
	
	The National Radio Astronomy Observatory is a facility of the National Science Foundation operated under cooperative agreement by Associated Universities, Inc.
	%%%%%%%%%%%%%%%%%%%%%%%%%%%%%%%%%%%%%%%%%%%%%%%%%%
	%==============================================================================================================
	
	
	%\newpage{}
	\bibliographystyle{mnras}
	\bibliography{_mnras} % if your bibtex file is called example.bib
	
	%\begin{thebibliography}{99}
	%bibitem[\protect\citeauthoryear{Author}{2012}]{Author2012}
	%\end{thebibliography}
	%%%%%%%%%%%%%%%%%%%%%%%%%%%%%%%%%%%%%%%%%%%%%%%%%%
	%==============================================================================================================
	
	%==============================================================================================================
	%%%%%%%%%%%%%%%%% APPENDICES %%%%%%%%%%%%%%%%%%%%%

	\appendix
	\section{Contribution of synchrotron radiation to the optical emission}
	\label{sec:appendix}
	
	Here we estimate the contamination of the optical emission from synchrotron radiation for the radio-detected quasars to assess whether this can explain the enhancement in the radio-detection of rQSOs.
	%	Some studies have proposed that the red colours of radio-selected quasars can be ascribed to excess emission at infrared wavelengths. The excess flux is suggested to either originate from synchrotron emission, particularly in FSRQs, and/or from stellar light in the host galaxy \citep[e.g.,][]{benn1998,francis2000,whiting2001}. 
	%	Here we address whether a red synchrotron contribution serves as a plausible explanation for the red colours of the rQSOs. 
	The SEDs of compact radio sources such as FSRQs (which forms part of the blazar AGN subclass) are well known for their ``double-humped" shape with a low-energy component extending from radio to UV/X-rays, and a high-energy component that extends from X-rays to GeV/TeV $\gamma$-rays. The synchrotron nature of the low-energy component is well established: FSRQs, for example, are well known for their low-synchrotron peaked SEDs with the peak frequency in the infrared regime ($< 10^{14}$\,Hz or wavelengths $>3 \mu$m). 
	Therefore, to investigate whether a red synchrotron component can explain the reddening of the rQSOs, we employed a model in which all the radio$-$optical power can be attributed to synchrotron emission, as in the case of FSRQs. 
	If the rQSOs are reddened by dominant synchrotron emission beamed along our line-of-sight, we would expect the optical synchrotron fluxes to be in agreement with the observed optical fluxes. 
	
\begin{figure}
		\centering
		\includegraphics[width=18pc]{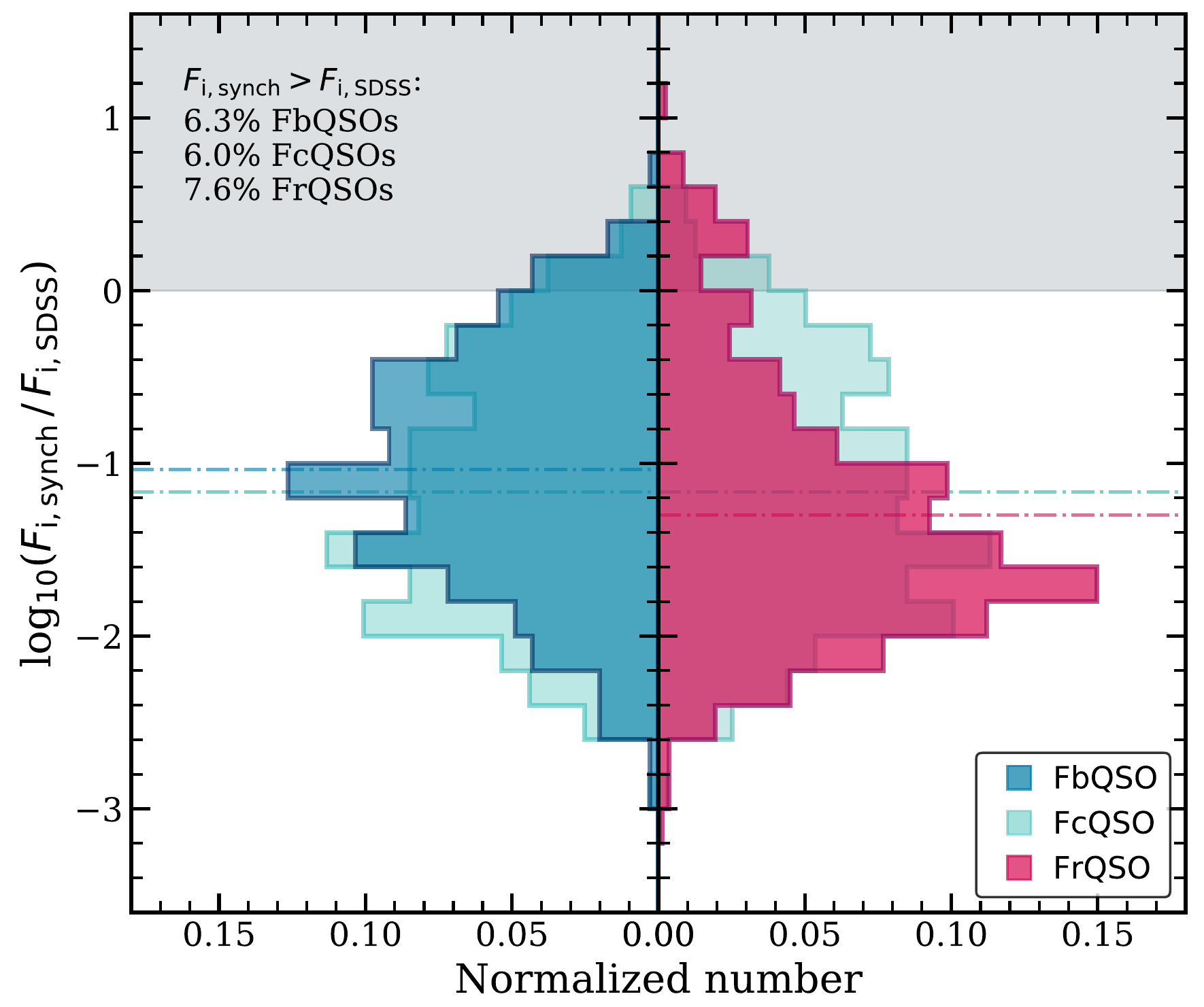}
		\caption{The ratio between the estimated synchrotron $i$-band flux and the observed SDSS $i$-band flux for the FbQSOs, FcQSOs and FrQSOs. The synchrotron $i$-band fluxes were estimated from the 1.4\,GHz flux using a synchrotron model of an FSRQ, 3FGL\,J0045.2-3704. The dashed lines indicate the median for each distribution, and the grey shaded area indicates where $F_{\rm i,synch} > F_{\rm i,synch}$. Based on this it is evident that synchrotron emission is not the main source of reddening of the rQSOs. }
		\label{fig:synchrotron iflux}
	\end{figure}

	The approach we have followed is to fit a synchrotron emission model \citep[e.g.,][]{finke2008,dermer2009} to the radio$-$optical data of 3FGL\,J0045.2-3704 \citep[see][]{klindt2017} to generate a prototypical template.
	We adopted the SED of 3FGL\,J0045.2-3704 because it has a blue quasar spectrum in the optical band, but also has a flat MIR-optical SED that is consistent with the observed properties of our quasars.
	We subsequently scaled this template to the 1.4\,GHz flux of our radio-detected quasars to estimate their expected $i$-band flux from a synchrotron component. 
	The ratios between the $i$-band fluxes estimated using the FSRQ synchrotron template ($F_{\rm i,synch}$) and the observed $i$-band SDSS fluxes ($F_{\rm i,SDSS}$) are shown in Figure~\ref{fig:synchrotron iflux} for the FIRST-detected bQSOs, cQSOs and rQSOs (FbQSOs, FcQSOs and FrQSOs, respectively). 
	%..............................................................................................................
	%..............................................................................................................
	
	For the majority of the quasars in all three samples a synchrotron model underpredicts the $i$-band fluxes by at least an order of magnitude.
	Only 6\% of FbQSOs and FcQSOs and 8\% of FrQSOs have $F_{\rm i,synch}$ greater than $F_{\rm i,SDSS}$.
	This suggests that the optical bands are not contaminated by synchrotron emission from the radio jet for the majority of the radio-detected quasars in our sample and, hence, this cannot be the primary source of reddening of the rQSOs. 
	Indeed, as can be seen in Figure~\ref{fig:synchrotron iflux}, the majority of the rQSOs will have an extremely small synchrotron contribution to the optical emission, which is a direct consequence of the pre-ponderance of low radio luminosities from the rQSOs. Consequently, the majority of rQSOs are expected to have a lower synchrotron contribution in the optical band than the bQSOs and cQSOs.
	
	%==============================================================================================================
	%==============================================================================================================
	
	% Don't change these lines
	\bsp	% typesetting comment
	\label{lastpage}
\end{document}